\documentclass[%
 aps,
 amsmath,amssymb,
 reprint,%
superscriptaddress,
longbibliography
]{revtex4-1}

\usepackage[hmargin=.55in,vmargin=0.75in]{geometry}
\usepackage{blkarray}
\usepackage{booktabs}

 \usepackage{tikz,pgfplots}
\usepackage[colorlinks=false,citecolor=blue,urlcolor=blue]{hyperref}
\usepackage{dcolumn}
\usepackage{bm}
\usepackage{tabu}
\usepackage{calc}
\usepackage{xspace}
\usepackage{mathtools}

\makeatletter
\newcommand*{\balancecolsandclearpage}{%
  \close@column@grid
  \cleardoublepage
  \twocolumngrid
}
\makeatother

\graphicspath{{figures/}}

\usepackage{animate}

\newcommand{\heading}[1]{\paragraph*{#1}}

\newcommand{\ket}[1]{\left|#1\right\rangle}

\newcommand{\bra}[1]{\left\langle#1\right|}

\newcommand{\G}[1]{\(\Gamma_{#1}\)}

\newcommand{\Ctv}{$C_{3v}$\xspace}

\newcommand{\unit}[1]{\hat{\mathbf{#1}}}

\newcommand{\semi}{;\,\,}

\definecolor{fgred}{rgb}{0.85 ,0 ,0}
\definecolor{fgorange}{rgb}{.9 ,0 ,.45}
\definecolor{fgblack}{rgb}{0 ,0 ,0}

\newcommand\T{\rule{0pt}{2.6ex}}       
\newcommand\B{\rule[-1.2ex]{0pt}{0pt}} 

\begin{document}
\title[]{Giant permanent dipole moment of 2D excitons bound to a single stacking fault}


%



\author{Todd~Karin}
\thanks{These authors contributed equally to this work.}
\affiliation{Department of Physics, University of Washington, Seattle, Washington 98195, USA}
\author{Xiayu Linpeng}
\thanks{These authors contributed equally to this work.}
\affiliation{Department of Physics, University of Washington, Seattle, Washington 98195, USA}
\author{M.M. Glazov}
\affiliation{Ioffe Institute, 194021 St.-Petersburg, Russia}
\author{M.V. Durnev}
\affiliation{Ioffe Institute, 194021 St.-Petersburg, Russia}
\author{E.L. Ivchenko}
\affiliation{Ioffe Institute, 194021 St.-Petersburg, Russia}
\author{Sarah~Harvey}
\affiliation{Department of Physics, University of Washington, Seattle, Washington 98195, USA}
\author{Ashish~K.~Rai}
\affiliation{Lehrstuhl f\"ur Angewandte Festk\"orperphysik, Ruhr-Universit\"at Bochum, D-44870 Bochum, Germany}
\author{Arne~Ludwig}
\affiliation{Lehrstuhl f\"ur Angewandte Festk\"orperphysik, Ruhr-Universit\"at Bochum, D-44870 Bochum, Germany}
\author{Andreas D. Wieck}
\affiliation{Lehrstuhl f\"ur Angewandte Festk\"orperphysik, Ruhr-Universit\"at Bochum, D-44870 Bochum, Germany}
\author{Kai-Mei C. Fu}
\affiliation{Department of Physics, University of Washington, Seattle, Washington 98195, USA}
\affiliation{Department of Electrical Engineering, University of Washington, Seattle, Washington 98195, USA}

\date{\today}

\begin{abstract}
We investigate the magneto-optical properties of excitons bound to single stacking faults in high-purity GaAs. We find that the two-dimensional stacking fault potential binds an exciton composed of an electron and a heavy-hole, and confirm a vanishing in-plane hole $g$-factor, consistent with the atomic-scale symmetry of the system. The unprecedented homogeneity of the stacking-fault potential leads to ultra-narrow photoluminescence emission lines (with full-width at half maximum ${\lesssim 80~\mu \text{eV} }$) and reveals a large magnetic non-reciprocity effect that originates from the magneto-Stark effect for mobile excitons. These measurements unambiguously determine the direction and magnitude of the giant electric dipole moment (${\gtrsim e \cdot 10~\text{nm}}$) of the stacking-fault exciton, making stacking faults a promising new platform to study interacting excitonic gases.
\end{abstract}

\maketitle


\heading{Introduction.}

The stacking fault (SF), a planar, atomically thin defect, is one of the most common extended defects in zinc-blende, wurtzite, and diamond semiconductors. A fundamental understanding of the SF potential is important for determining how the defect affects semiconductor device performance~\cite{Guha1993,Colli2003}, engineering heterostructures based on crystal phase~\cite{Caroff2011,Akopian2010,Assali2013}, and providing a new two-dimensional (2D) platform for fundamental physics~\cite{Butov2002,High2009}.
Here we report on excitons bound to large-area, single SFs in high-purity GaAs, a unique system where SFs are easily isolated with far-field optical techniques. The atomic smoothness of the potential and extreme perfection of the surrounding semiconductor result in ultra-high optical homogeneity (${\lesssim80~\mu\text{eV}}$).  This enables optical resolution of the SF exciton fine-structure and thus direct measurement of the giant built-in dipole moment (${\gtrsim e \cdot 10~\text{nm}}$) via the magneto-Stark effect. These results indicate that the extremely-homogeneous SF potential may be promising for studies of many-body excitonic physics, including coherent phenomena~\cite{Snoke2002,Butov2002bec,Shilo2013}, spin currents~\cite{High2013}, superfluidity~\cite{Fogler2014}, long-range order~\cite{Gorbunov2006,Snoke2011,Nelson2009,Balili2009,High2012}, and large optical nonlinearities~\cite{Amo2010,Nguyen2013,Kammann2012}.


\begin{figure*}[hbt]
\includegraphics{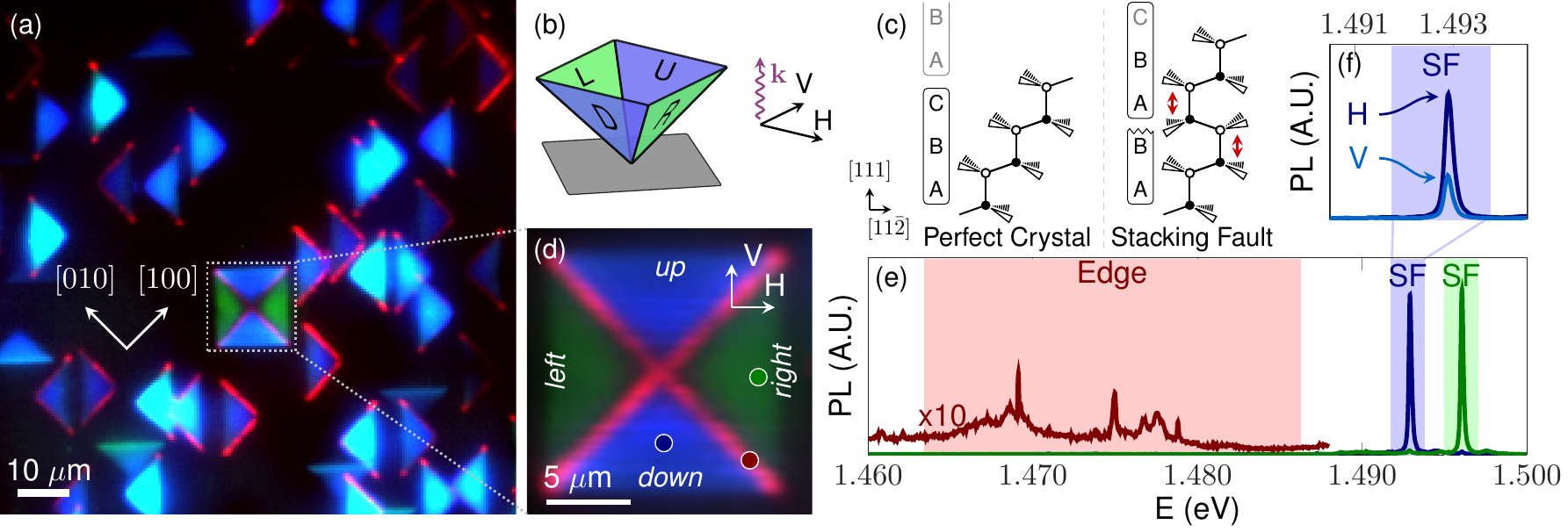}
\caption{
(a) Confocal scan of SF structures. The image is formed by coloring emission in different wavelength bands as red, blue or green, as depicted in e. Excitation at 1.53~eV, $100~\mu$W, 1.9~K, excite and collect H polarization (see b).
(b) Diagram of SF pyramid. The \emph{up, down, left} and \emph{right} SFs are labeled, along with the H and V polarizations.
(c) Comparison of perfect zinc-blende and stacking fault crystal structure.
(d) Detail of SF pyramid structure. Excitation at 1.53~eV, $100~\mu$W, 1.7~K. 
(e) Low power PL spectra at colored dots in d. Polarizations: blue - excite/collect H; green, red - excite/collect V. Broad-band luminescence is observed from the SF edges (red). 1.53~eV, 2~\(\mu\)W, 1.7~K. 
(f) PL from \emph{down} SF (blue dot in d). Polarizations: dark blue - excite/collect H; light blue - excite/collect V.
}
\label{fig:confocal}
\end{figure*}

\heading{Stacking fault photoluminescence.}
Figure~\ref{fig:confocal}(a) shows a spectrally resolved confocal scan of SF structures in a GaAs epilayer, excited with an above band-gap laser ($1.65$~eV, 1.5~K)~\cite{sfsupplement}. 
The image is colored red, green or blue according to three characteristic emission bands shown in Fig.~\ref{fig:confocal}e. The narrow-band PL at 1.493 and 1.496~eV originates from excitons, electron-hole pairs, bound to the 2D SF potential~\cite{Kasai1998,Lahnemann2014}.
The sample consists of a 10~$\mathrm{\mu m}$ GaAs layer on 100 nm AlAs on a 5~nm/5~nm AlAs/GaAs (10$\times$) superlattice grown directly on a semi-insulating (100) GaAs substrate. Stacking fault structures nucleate near the substrate-epilayer interface during epitaxial growth~\cite{sfsupplement}.

The physical origin of the potential can be understood from the atomic structure of the SF defect: the lattice-plane ordering in the [111] direction of zinc-blende is modified by subtracting a layer (intrinsic SF, see Fig.~\ref{fig:confocal}c) or adding a layer (extrinsic SF).
The intrinsic SF can be viewed as a monolayer of wurtzite (AB\,AB stacking) surrounded by zinc-blende (ABC\,ABC stacking)~\cite{Algra2008,Caroff2011}. Due to the band offset~\cite{Belabbes2012,Spirkoska2009,Heiss2011} and spontaneous polarization at the stacking fault~\cite{Lahnemann2012}, electrons and/or holes are attracted to the SF plane. While useful for physical motivation, this bulk phase change model must be taken with caution when applied to atomically thin SFs, which can deviate from simple theory~\cite{Vainorius2015}. 
Here, however, we find that single SFs in bulk GaAs bind excitons, confirming that the potential is attractive for at least one carrier.

In the confocal scan in Fig.~\ref{fig:confocal}(a), most of the SF defects appear as single triangles, which we identify as a pair of nearby SFs~\cite{Fung1997,Wang2000}. Because the binding energy of excitons to a pair of SFs depends on the distance between the SFs~\cite{Corfdir2012}, the PL emission energy from excitons bound to these structures has a high variability of \(10\)~meV between structures. Strikingly, this inhomogeneity disappears when four SFs grow in an inverted pyramid structure consisting of four well-isolated $\{111\}$ SF planes~[Fig.~\ref{fig:confocal}(b)], which we refer to as \emph{up}, \emph{down}, \emph{left} and \emph{right} \cite{Kakibayashi1984}. The full width at half-maximum (FWHM) of the SF PL line in our sample is $(77\pm19)~\mu$eV at zero magnetic field~\cite{sfsupplement}, somewhat narrower than excitonic lines associated with stacking faults in previous work~\cite{Kasai1998,Komatsu1988}.   In comparison, the narrowest reported linewidth for a GaAs/AlGaAs quantum well is 130~\(\mu\)eV~\cite{Poltavtsev2014}, while PL linewidths from analogous zinc-blende/wurtzite quantum discs in nanowires range from \(0.6-10~\)meV~\cite{Heiss2011,Graham2013,Pal2008,Signorello2014}. This unprecedented homogeneity allows us to resolve the SF-bound exciton fine structure

\heading{Nature of hole in SF exciton.}

\begin{figure}[hbt]
\includegraphics{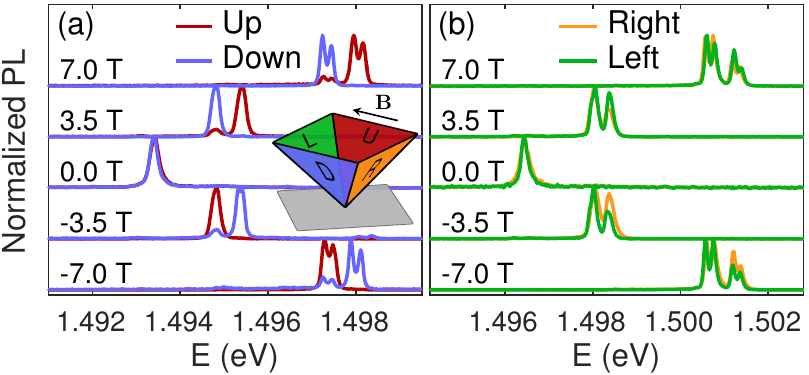}
\caption{
(a) Spectra from \emph{up} and \emph{down} SFs as a function of in-plane magnetic field. The spectra show a non-reciprocity with applied magnetic field. Excite at 1.65~eV, 0.5~\(\mu\)W, 1.6~K, excite and collect H. Inset shows the geometry of the stacking fault pyramid and applied magnetic field.
(b) Spectra from \emph{left} and \emph{right} SFs as a function of partially out-of-plane magnetic field. The spectra are similar at positive and negative fields. Excitation at 1.65~eV, 0.5~\(\mu\)W, 1.6~K, excite and collect V. 
}
\label{fig:spectra}
\end{figure}

Experimentally, we determine that the SF exciton is composed of an electron and a heavy-hole using polarization resolved PL, consistent with the atomic-scale symmetry of the system~\cite{sfsupplement}. For linearly polarized light incident from above (along the [001] axis), the largest overlap between the light polarization and the in-SF-plane heavy-hole dipole occurs when exciting and collecting along the H direction for the \emph{down} SF [Fig.~\ref{fig:confocal}(d)], in agreement with our experimental data [Fig.~\ref{fig:confocal}(f)]. On the other hand, the main dipole moment for the light-hole exciton is along the SF normal, which would give rise to a maximum signal at V polarization, contrary to what is observed.
Further, we also note that no hole Zeeman splitting is observed for in-plane magnetic fields $B$ up to 7~T (Fig.~\ref{fig:spectra}). This observation is fully consistent with our symmetry analysis, which finds that $B$-linear splitting for in-plane fields is forbidden for heavy-holes but allowed for light-holes~\cite{sfsupplement}. The substantial separation of the heavy- and light-hole states prevents their magnetic-field induced mixing, in line with experiments on GaAs nanowires~\cite{Spirkoska2012,sfsupplement}.



\heading{Non-reciprocal photoluminescence.}
PL from SFs shows a remarkable non-reciprocity with in-plane applied magnetic field: Figure~\ref{fig:spectra}(a) shows that the PL detected in linear polarization from the \emph{up} SF occurs at a different energy depending on whether the magnetic field is parallel (positive) or antiparallel (negative) to the $[\bar 110]$ axis. Interestingly, the \emph{down} SF demonstrates the opposite behavior. Such an asymmetric behavior of the PL is surprising because in general, time reversal symmetry makes $\mathbf B$ and $-\mathbf B$ equivalent~\footnote{Our sample is non-magnetic and we use linearly polarized light to avoid dynamic polarization of nuclear spins.}. The observed non-reciprocal behavior of the PL spectrum with respect to inversion ${\mathbf B\to -\mathbf B}$ is only possible if the PL arises from moving excitons. In this case, time reversal changes the direction of both the magnetic field and the exciton wavevector $\mathbf K$.


Based on the \Ctv point symmetry of the SF and time reversal invariance, the effective Hamiltonian for an exciton moving in the presence of an in-SF-plane magnetic field $\mathbf B$ is
\begin{equation}\label{eq:KB}
\mathcal H_{KB} = \frac{g_e}{2}\mu_B (\sigma_x B_x + \sigma_y B_y)  + \beta B^2  + \beta' [\mathbf K \times \mathbf B]_z,
\end{equation}
where $g_e$ is the electron $g$-factor, $\mu_B$ is the Bohr magneton, $\sigma_{x,y}$ are the electron spin Pauli matrices, $\beta$ is a parameter describing the excitonic diamagnetic shift, and $\beta'$ is a constant responsible for the non-reciprocal effect~\cite{sfsupplement}. In Eq.~\eqref{eq:KB} we only retain 1st- and 2nd-order terms in $\mathbf B$ and
use a frame of axes related to the SF plane: ${z\parallel[111]}$ is the SF normal, ${x||[11\bar 2]}$ and ${y||[\bar 110]}$.
Each symmetry-derived term in Eq.~\eqref{eq:KB} 
manifests itself in the energetic shift of the SF PL lines with magnetic field (Fig.~\ref{fig:spectra}). The first term is the electron Zeeman effect and gives rise to the doublets visible at \(\pm\)7~T, since an electron with a particular spin projection can recombine with the corresponding hole. The second term is the exciton diamagnetic shift, arising from the magnetic-field-induced shrinking of the exciton wavefunction~\cite{Knox}. The last term is the magneto-Stark effect, which, as we show below, quantitatively explains the non-reciprocal PL spectra.



\begin{figure*}[hbt]
\includegraphics[]{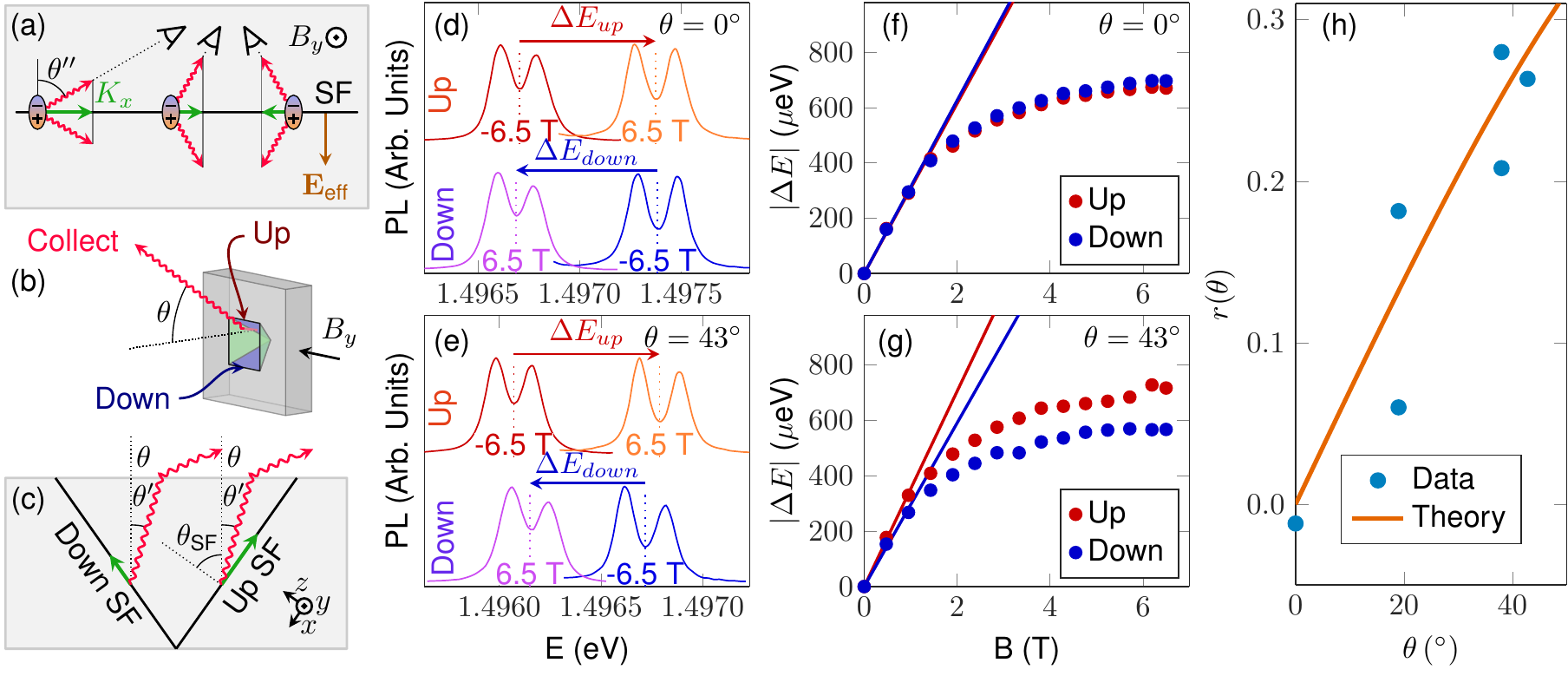}
\caption{
(a) Because of conservation of in plane momentum during exciton recombination, the angle of light emission depends on the exciton wavevector. Collecting different angles probes different exciton momenta. The SF has a built in potential that creates a zero-field dipole moment for the SF exciton. In the exciton frame of reference, the in-plane magnetic field becomes an out-of-plane electric field, leading to the magneto-Stark effect.
(b) Spectra of \emph{up} and \emph{down} SFs as a function of \(\theta\) and in-plane magnetic field \(B_y\).
(c) Light from the \emph{up} SF originates from excitons with larger \(K_x\) than light from the \emph{down} SF (for \(\theta>0\)).
(d-e) Spectra of \emph{up} and \emph{down} SF at positive and negative \(B_y\) for \({\theta=0^\circ}\) and \({43^\circ}\).  At \({\theta=0^\circ}\), \(\Delta E_{up}\) and  \(\Delta E_{down}\) have the same magnitude, while for \({\theta=43^\circ}\), the magnitude of \(\Delta E_{up}\) is larger than \(\Delta E_{down}\).
(f-g) Splitting \(\Delta E_{up/down}\) as a function of magnetic field. Data are obtained from Voigt fits to spectra similar to those shown in d-e. Solid lines are a fit to \({\Delta E = \mbox{\textsl{a}} B}\) for the first three data points.
(h) The ratio of \({B=0}\) slopes, Eq.~\eqref{eq:r}, depends only on geometrical constraints. The theory (solid line) has no adjustable parameters. Data for other angles in~\cite{sfsupplement}.
}
\label{fig:angle}
\end{figure*}

The experimental geometry, Fig.~\ref{fig:confocal}(b), is such that only light emitted normal to the sample surface is collected.
For a high quality 2D potential, in-plane exciton momentum is transferred to the photon during recombination, as depicted in Fig.~\ref{fig:angle}(a). This conservation of momentum implies
\begin{equation}
\label{eq:anglekagree} 
K_x = \frac{\omega n}{c} \sin \theta'',
\end{equation}
where \(\theta''\) is the angle between the SF normal and the emitted photon momentum inside the semiconductor, Fig.~\ref{fig:angle}(a), $\omega$ is the photon frequency, $n$ is the refractive index and $c$ the speed of light. Thus, the collected SF PL arises only from excitons with a specific center of mass momentum~\footnote{Experimentally, the \({\text{NA}=0.7}\) objective lens collects luminescence from a range of angles. In our system, light is collected from excitons momenta within 7\% of \(\hbar K_x\) in Eq.~\eqref{eq:anglekagree}.}. The last term in Eq.~\eqref{eq:KB} provides, for a fixed $K_x$ (Eq.~\ref{eq:anglekagree}), an odd in \(B_y\) contribution to the overall PL energy shift, giving rise to a magnetic non-reciprocity effect. It is worth noting that the \emph{up} and \emph{down} SFs are related by a mirror reflection in the (110) plane and such a reflection is accompanied by ${B_y\to -B_y}$, resulting in the opposite behavior of \emph{up} and \emph{down} PL spectra observed in Fig.~\ref{fig:spectra}(a).


\heading{Magneto-Stark effect.} The physical origin of the non-reciprocal PL is the magneto-Stark effect, the interaction of a moving exciton's electric dipole moment with a magnetic field~\cite{Gross1961,Thomas1961}. The effect can be understood with a relativistic argument: motion with velocity ${ \mathbf v = (\hbar K_x/M) \unit{x}}$ through a magnetic field \({\mathbf B = B_y \unit y}\) gives rise to an electric field ${\mathbf E_{\rm eff} = \hbar K_x B_y/(Mc) \unit{z}}$ in the moving frame of reference, where $M$ is the exciton mass in translational motion and $c$ the speed of light. Since for the SF, ${\unit{z} \propto [111]}$ and $-\unit{z}$ directions are not equivalent, the SF-bound exciton has a non-zero dipole moment \({\mathbf p=ed_{he} \unit z}\), where ${e=|e|}$ is the elementary charge, and $d_{he}$ is the average separation between the hole and electron along the $z$-axis. 
The Stark effect \({H_s=-\mathbf p\cdot \mathbf E_{\rm eff}}\) in the exciton's reference frame thus becomes the magneto-Stark effect:
\begin{equation}\label{eq:magnetostark}
\mathcal H_\text{S} = - \frac{e\hbar}{Mc} d_{he} K_x B_y,
\end{equation}
in agreement with Eq.~\eqref{eq:KB} with ${\beta'=-e\hbar d_{he}/(Mc)}$, see Ref.~\cite{Knox,sfsupplement} for formal derivation.

Physically, the dipole moment of a SF bound exciton is a consequence of symmetry breaking and spontaneous polarization similar to that in zinc-blende/wurtzite heterostructures~\cite{Lahnemann2014,Jahn2012}. The hole in the exciton is presumably localized in the SF plane while the electron is weakly bound via the Coulomb interaction. The spontaneous polarization shifts the electron cloud to one side of the SF, resulting in a giant excitonic dipole moment.



Equations~\eqref{eq:KB}-\eqref{eq:magnetostark} predict that the asymmetric energy shift of exciton PL is linearly related to the in-plane wavevector $\mathbf K$. Since the angle of light collection determines the exciton momentum [Eq.~\eqref{eq:anglekagree}], we test the applicability of the model by recording spectra of the \emph{up} and \emph{down} SFs as a function of the collection angle \(\theta\) and magnetic field \(B_y\) [Fig.~\ref{fig:angle}(b)]. The collection angle is related to the emission angle from the up/down SF by \({\sin \theta  =n \sin\theta' = \pm n \sin (\theta'' - \theta_\text{SF})  }\), where \(\theta_\text{SF}\) is the angle the SF normal $\unit z|| [111]$ makes with [001] [Fig.~\ref{fig:angle}(c)].


In this experiment, we modified the collection angle by mounting the sample at different angles. Since the sample was removed from the cryostat to change the angle, different SF pyramids were used at different angles. This does not introduce artifacts because of the extreme similarity of different SFs, which have a standard deviation of line-center energies of only 57~$\mu$eV, less than the linewidth. Spectra were acquired with $B_y$ ranging from $-6.5$~T to 6.5~T on the \emph{up} and \emph{down} SFs. We fit the spectra to one or a sum of two Voigt function(s) depending on whether the electron Zeeman splitting is resolved. The singlet or doublet line center is denoted \(E_{up/down}(B_{y})\). The part of the exciton energy odd with magnetic field is found by computing 
\begin{equation}
\Delta E_{up/down}(B_{y}) = E_{up/down}(B_{y}) - E_{up/down}(-B_{y}) 
\end{equation}
It follows from Eq.~\eqref{eq:magnetostark} that the asymmetric shift is 
\begin{equation}\label{eq:props}
\Delta E_{up/down}(B_{y}) = \mp 2 n \hbar \omega \frac{e d_{he}}{M c} \sin( \theta_{SF} \pm \theta' ) B_y .
\end{equation}
Thus the proportionality constant of \(\Delta E_{up/down}\) vs. \(B_y\) provides a measurement of the SF exciton's built-in dipole moment. The experimental values and first-order theory for \(\Delta E\) are shown in Fig.~\ref{fig:angle}(f)-(g).
Further, the ratio
\begin{equation}\label{eq:r}
r(\theta) = \frac{ |\Delta E_{up}| - |\Delta E_{down}| }{ \frac{1}{2} \left( |\Delta E_{up}| + |\Delta E_{down} |\right) }
\end{equation}
depends (to first order in $B_y$) only on the experimental geometry and the index of refraction: $r(\theta)$ vanishes for collection angle ${\theta=0}$ and increases as a function of \(\theta\) [Fig.~\ref{fig:angle}(h)]. We obtain good agreement between \(r(\theta)\) calculated experimentally from the \({B=0}\) slope of \(\Delta E\) without any fit parameters [Fig.~\ref{fig:angle}(h)].


Further, by fitting $\Delta E_{up/down}(B_y)$ with a $B_y$-linear function, we can estimate the dipole moment of the exciton ${p = e d_{he} = e\cdot (10^{+20}_{-1})~\text{nm}}$. The main uncertainties result from the accuracy of the \(B_y\)-linear fit and the value of the in-\((111)\)-plane heavy-hole mass, which depends on the details of the SF potential~\cite{sfsupplement}.
The exciton mass can be roughly estimated as $0.17\,m_o$, the sum of the bulk-GaAs in-\((111)\)-plane heavy-hole mass and the isotropic electron mass, where $m_o$ is the free electron mass. In addition, we note the magneto-Stark induced splitting saturates at high fields [Fig.~\ref{fig:angle}(f,g)], possibly due to a decreased exciton dipole moment from the magnetic-field-induced shrinking of the exciton wavefunction. Future work will investigate exciton confinement potentials consistent with the observed dipole moment, diamagnetic shift and saturation of the magneto-Stark effect. A microscopic understanding of the confinement potential may enable predictions for the binding potential and excitonic dipole moment for SFs in other semiconductors.




\heading{Conclusion.} We have shown that SFs in GaAs are an almost perfect 2D potential which binds heavy-hole excitons. These excitons freely propagate in the SF plane, a conclusion confirmed via the magneto-Stark effect. Further, an asymmetry of the SF potential induces a giant dipole moment of the SF-bound exciton. Such excitons could be useful for studying the many-body physics of interacting dipoles. In conventional excitonic systems, typical electron-hole separations are on the order of several~nm~\cite{Butov2002,Warburton2002}, whereas the SF-bound exciton has a gigantic electron-hole separation of 10~nm and the possibility to modify this value with an applied field. In addition, the ultra-narrow linewidths in the SF system will allow the small energy shifts present in many-body interactions to be observed. As a rough estimate, the interaction energy of two such dipoles will exceed the SF FWHM of 77~$\mu$eV when the exciton density is greater than 230~${\mu\text{m}^{-2}}$. Using a wavefunction size of approximately 10~nm, the critical density for exciton overlap in the 2D potential is 10\,000~${\mu\text{m}^{-2}}$. Therefore, the SF-bound exciton system could show sizable dipole-dipole interactions and may demonstrate coherent phenomena at reasonable exciton densities.


\heading{Acknowledgements.} This material is based upon work supported by the National Science Foundation under Grant Number 1150647 and the National Science Foundation Graduate Research Fellowship under grant number DGE-1256082, and in part by the State of Washington through the University of Washington Clean Energy Institute. The Ioffe team has been partially supported by RFBR, RF President grant No. MD-5726.2015.2 and Dynasty foundation. A.K.R., A.L., and A.D.W. acknowledge partial support of Mercur Pr-2013-0001, DFG-TRR160, BMBF - Q.com-H  16KIS0109, and the DFH/UFA  CDFA-05-06.
We would like to acknowledge helpful discussions with John Schaibley, Pasqual Rivera, Xiaodong Xu, David Cobden and Matt McCluskey.

\balancecolsandclearpage

\renewcommand{\thesection}{S\arabic{section}}   

\renewcommand{\theequation}{S\arabic{equation}}
\renewcommand{\thefigure}{S\arabic{figure}}

\widetext
\begin{center}
\textbf{\large Supplemental Materials: Fundamental properties of 2D excitons bound to single stacking faults in GaAs}
\end{center}
\twocolumngrid

\section{Stacking Fault formation}\label{sec:growth}

\begin{figure}[hb]
\includegraphics{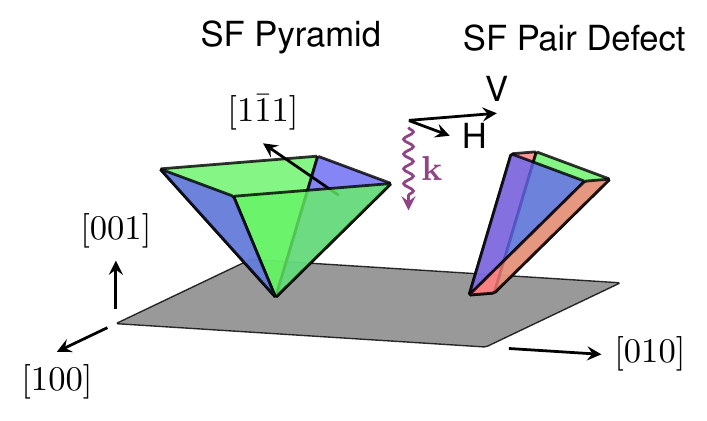}
\caption{Diagram showing the types of defects visible in the sample. The SF pyramid consists of four stacking faults arranged in a pyramid shape. The SF pair defect is a set of two SFs with a small \(\sim\)nm separation.}
\label{fig:sftypes}
\end{figure}

Stacking fault (SF) structures can grow from the substrate-epilayer interface during epitaxial growth~\cite{Kakibayashi1984,Chai1985}. In the present work, SFs form in a 10$~\mathrm{\mu m}$ GaAs layer grown by molecular beam epitaxy with room temperature electron density ${n \sim 1.9\times10^{14}~\mathrm{cm}^{-3}}$ and mobility ${\sim 7400~\text{cm}^2/\text{Vs}}$.  The entire structure consists of the 10~$\mathrm{\mu m}$ GaAs layer on 100 nm AlAs on a 5~nm/5~nm AlAs/GaAs (10$\times$)
 superlattice grown directly on a semi-insulating (100) GaAs vertical gradient freeze substrate (Wafer Technology Ltd), started with AlAs.  The sample was grown at a pyrometer temperature of 600$^\circ$C with the relatively low As$_4$ beam equivalent pressure of ${8 \times 10^{-6}~\text{Torr}}$, measured by a flux tube.  The growth rate was 0.7~ML/s for the GaAs and 0.35~ML/s for the AlAs. Oxide removal before growth was performed at 620$^\circ$C under As flux. This growth procedure resulted in oval defects in the sample surface, a feature commonly associated with stacking faults~\cite{Kasai1998,Kakibayashi1984,Haverkort1992,Chai1985}. We observe two types of stacking fault defects, a SF pyramid and a SF pair defect, shown in Fig.~\ref{fig:sftypes}. 
The size of the pyramid structure (14.1 $\mu$m top edge in Fig.~\ref{fig:confocal}a) is consistent with stacking faults that nucleate near the substrate-epilayer interface and grow along $\{111\}$ planes through the 10~\(\mu\)m thick epilayer.

\section{Photoluminescence spectroscopy system}\label{sec:methods}

The sample is excited with a continuous wave Sirah Matisse Ti:Sapphire laser. The laser is focused to a spot size of $\sim$1~$\mu$m on the sample using an aspheric lens (numerical aperture $0.77$) mounted inside a liquid helium immersion cryostat (Janis). Coarse positioning of the sample was performed with slip-stick positioners (Attocube). For confocal scanning, spatial selectivity is achieved with a pinhole in the intermediate image plane and a scanning mirror to raster the excitation and collection spot over the sample. The photoluminescence (PL) is imaged on a spectrometer (Andor).

\section{Stacking fault photoluminescence linewidth}\label{sec:deconvolution}

\begin{figure}[htb]
\includegraphics{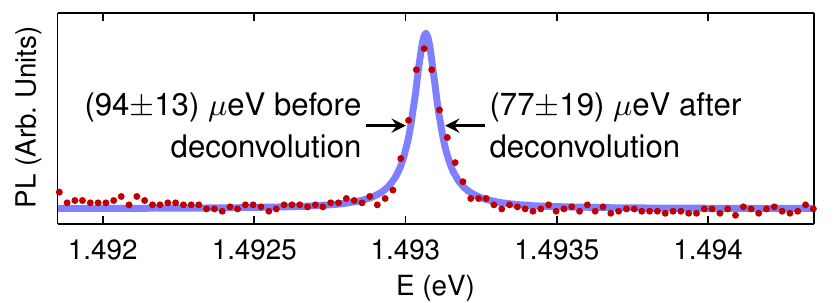} 
\caption{
High resolution PL spectrum of \emph{up} SF. The FWHM of the line is \( (77 \pm 19)~\mu\)eV from a weighted Lorentzian fit, taking into account the instrument resolution. Before deconvolution the FWHM is  $(94 \pm13)~\mu$eV. Excitation at 1.65~eV, 1.44~K, 1.4~nW.
}
\label{fig:77mueV}
\end{figure}

\begin{figure*}[htb]
\includegraphics{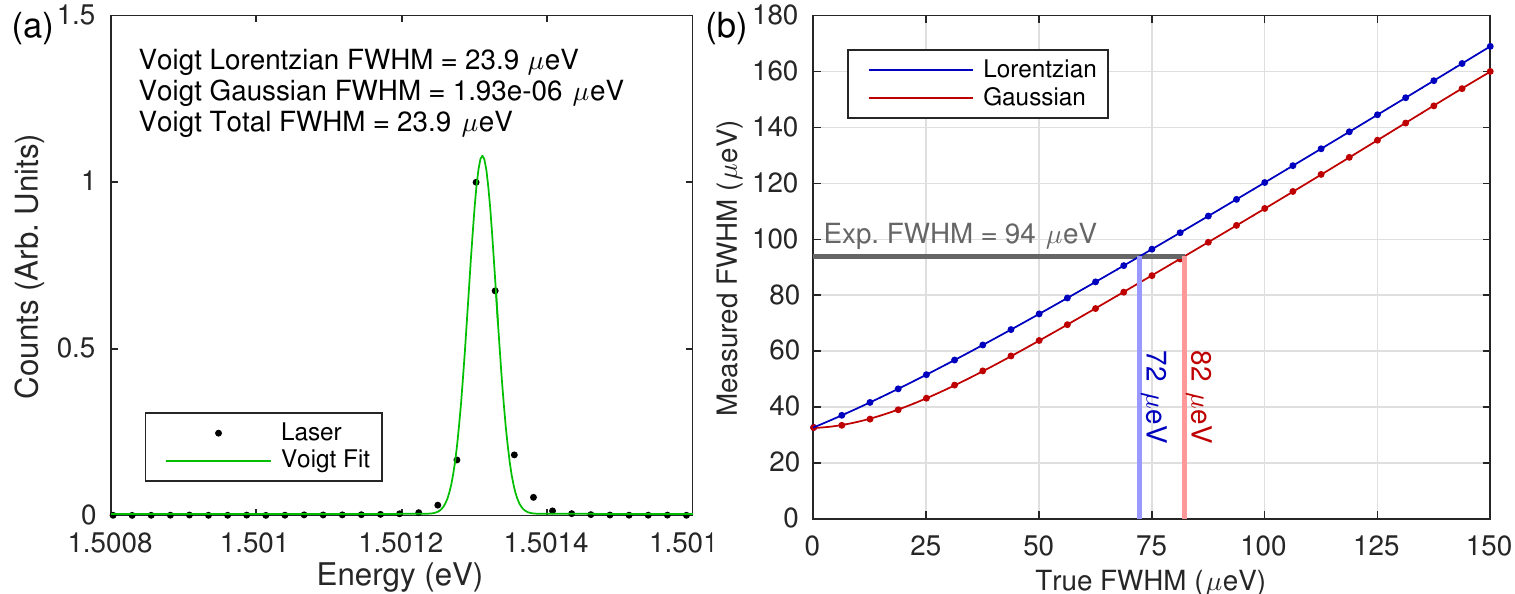}
\caption{ (a) Spectrum of the narrow band Ti:Sapphire laser used to determine the spectrometer instrument resolution. The best fit is a Voigt function with a $32.7~\mu$eV FWHM. The Voigt lineshape is the convolution of a Lorentzian and a Gaussian with best fit widths provided as an inset in the figure. (b) Convolution of the spectrometer instrument response in a with a Lorentzian or Gaussian lineshape. By interpolating backwards, the deconvoluted FWHM of a spectral line can be found.}
\label{fig:laserdeconvolution}
\end{figure*}

Figure~\ref{fig:77mueV} shows a high resolution PL spectrum of the SF. In order to extract the true PL linewidth, we need to take into account the spectral resolution of our setup. The spectrometer instrument resolution is found by taking a spectrograph of a narrow band Ti:Sapphire laser and fitting to a Voigt function, see Fig.~\ref{fig:laserdeconvolution}(a). We find that neither a Gaussian nor a Lorentzian accurately describe the spectral point spread function, so we use a Voigt fit. For spectral lines that are nearly as narrow as the spectrometer FWHM (full width at half maximum), the measured FWHM will be wider than the true FWHM. Figure~\ref{fig:laserdeconvolution}(b) shows the FWHM of a line obtained by the convolution of a Lorentzian or Gaussian spectral lineshape with the spectrometer response function. For example, a measured linewidth of 94~\(\mu\)eV corresponds to a true linewidth of 72 or 82~\(\mu\)eV, depending on whether the true lineshape is assumed to be Lorentzian or Gaussian. Hence, to evaluate the intrinsic linewidth of the SF emission we fit the spectrum in Fig.~\ref{fig:77mueV} with a weighted Lorentzian and use the deconvolution procedure [Fig.~\ref{fig:laserdeconvolution}(b)] to obtain an intrinsic PL linewidth of only \( (77 \pm 19)~\mu\)eV. Here, the uncertainty combines the original fit uncertainty and the uncertainty of the deconvolution procedure.

\section{Magneto-Stark Hamiltonian}\label{sec:magnetoStark}

Microscopically, the magneto-Stark effect can be derived from the Hamiltonian for an electron and hole in a magnetic field \(\bm B = B_y \unit y\):
\begin{equation}\label{eq:bHami}
H = \frac{\hbar^2}{2m_e} \left(\mathbf k_e + \frac{e}{\hbar c} \mathbf A \right)^2 + \frac{\hbar^2}{2m_h} \left(\mathbf k_h - \frac{e}{\hbar c} \mathbf A \right)^2 ,
\end{equation}
where \(\mathbf{k}_e\) (\(\mathbf{k}_h\)) is the wavevector of the electron (hole), \(m_e\) (\(m_h\)) is the effective mass of the electron (hole) and the vector potential is \({\mathbf A = B_y z \, \unit x}\). The Coulomb interaction between the electron and the hole as well as the SF potential are omitted in Eq.~\eqref{eq:bHami} for brevity.
We use the standard definition of center of mass (COM) and relative coordinates
\begin{equation}
\begin{aligned}
\mathbf k_e &= \frac{m_e}{M} \mathbf K + \mathbf k \\
\mathbf k_h &= \frac{m_h}{M} \mathbf K - \mathbf k \\
\end{aligned}
\end{equation}
where \(M = m_e +m_h\). With these substitutions, Eq.~\ref{eq:bHami} becomes
\begin{equation}
\begin{aligned}
H &= \overbrace{\frac{\hbar^2 K^2}{2M}}^\text{COM Kinetic} + \overbrace{\frac{\hbar^2 k^2}{2 \mu}}^\text{Rel. Kinetic}  \overbrace{- \frac{e\hbar}{Mc} (z_h-z_e) K_x B_y}^\text{COM Magneto-Stark}  \\
& + \overbrace{\frac{e\hbar}{c} \left(\frac{z_h}{m_h}+ \frac{z_e}{m_e} \right) k_x B_y}^\text{Orb. Zeeman} + \overbrace{ \frac{e^2 B_y^2}{2  c^2} \left( \frac{z_e^2}{m_e} + \frac{z_h^2}{m_h} \right) }^\text{Diamagnetic}
\end{aligned}
\end{equation}
where  \( {\mu =  (m_e^{-1} + m_h^{-1})^{-1} }\)~\cite{Knox}. The first term is the COM kinetic energy of the exciton. The second term is the kinetic energy associated with the electron-hole relative motion. The third term describes the magneto-Stark effect for the exciton COM motion. The fourth term describes the orbital Zeeman effect~\cite{Knox}. The fifth term is the harmonic potential created by the magnetic field which produces the diamagnetic shift and would produce Landau quantization for higher magnetic fields. The COM magneto-Stark term is the same as Eq.~\eqref{eq:magnetostark} in the main text, derived from relativistic arguments.


\section{Heavy hole -- light hole splitting}\label{sec:grouptheory}

The possible carrier spin states entering into the SF exciton can be predicted on general symmetry considerations. Due to spin orbit coupling, the valence band in bulk GaAs splits into the heavy-hole/light-hole (\({j=\frac{3}{2}}\)) bands and split-off (\({j=\frac{1}{2}}\)) band. At the \(\Gamma\) point, the heavy-hole (HH) and light-hole (LH) bands are degenerate and transform according to the four-dimensional \(\Gamma_8\) irreducible spinor representation of the \(T_d\) point symmetry group~\cite{BirPikus}, see Refs.~\cite{Koster1963,Dresselhaus} for notations.

\begin{figure}[tb]
\includegraphics{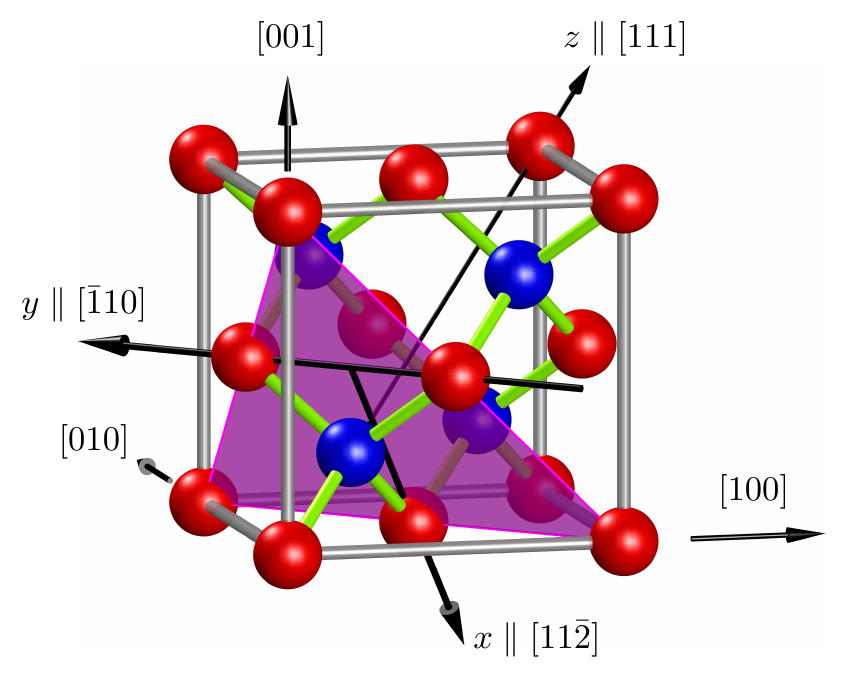} 
\caption{GaAs crystalline lattice. Stacking fault plane is a (111) plane, shown as a triangle. The mirror reflection plane for a SF sends \(y \to -y\). Note that \(x \to -x\) is not a mirror reflection plane.}
\label{fig:sym}
\end{figure}

\begin{figure*}[t]
\includegraphics{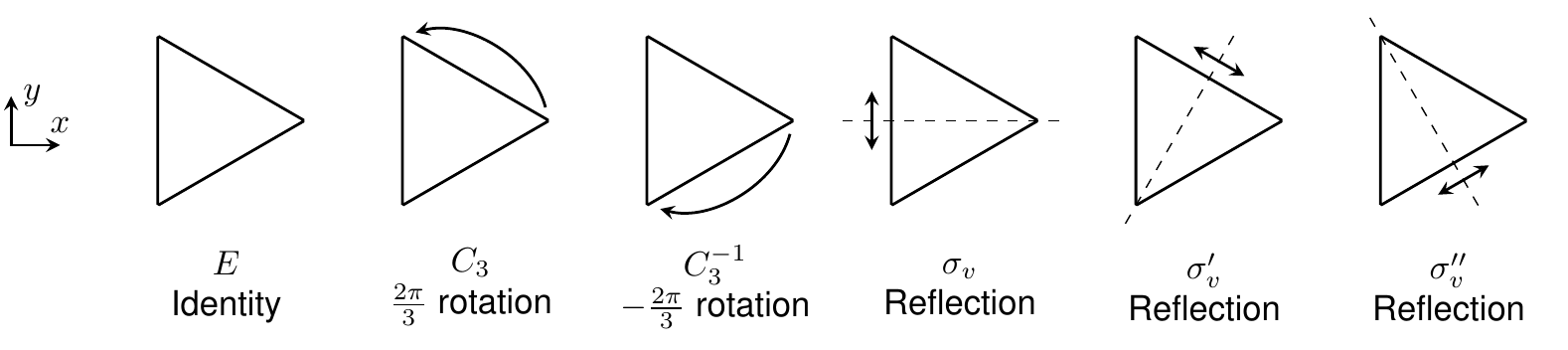}
\caption{Symmetry operations of $C_{3v}$, the point group for a (111) SF in GaAs.} 
\label{fig:c3vOps}
\end{figure*}

A SF oriented in one of the $\{111\}$ planes possesses the lower \(C_{3v}\) point symmetry, as illustrated in Fig.~\ref{fig:sym}. 
The \(C_{3v}\) symmetry group contains six symmetry operations: E (identity), \(C_3\) (rotation by $2\pi/3$ about $\unit z$), \(C_3^{-1}\) (rotation by $-2\pi/3$ about $\unit z$), $\sigma_v$ (reflection \(y\to-y\)),  $\sigma_v'$ ($\sigma_v'=\sigma_v C_3^{-1}$), and  $\sigma_v''$ ($\sigma_v''=\sigma_v C_3$). These operations are depicted in Fig.~\ref{fig:c3vOps}.
We note that the SF symmetry group is different from the $C_{6v}$ symmetry of wurtzite~\cite{Birman1959}.

When a SF is introduced into a zinc-blende crystal, the degeneracy of the valence and conduction band edges could be lifted. The compatibility analysis shows that the conduction band edge transforms according to the two-dimensional irreducible representation $\Gamma_4$ of the C$_{3v}$ point group, i.e. the conduction band does not split. On the other hand, the degeneracy of the valence band is lifted: the $\Gamma_8$ irreducible representation of T$_d$ decomposes into ${\Gamma_4 \oplus \Gamma_5 \oplus \Gamma_6}$ of C$_{3v}$. The two degenerate states $\Gamma_4$ transform as the spinors 
${|3/2,1/2\rangle}, {|3/2,-1/2\rangle}$ (or ${|1/2,1/2\rangle}, {|1/2,-1/2\rangle}$), where $|j,m_j\rangle$ signifies the basic function for angular momentum $j$ and angular momentum $z$-component $j_m$. The $\Gamma_5$ and $\Gamma_6$ hole states transform as linear combinations the spinors ${|3/2,3/2\rangle}$ and ${|3/2,-3/2\rangle}$, see Table I. At ${{\bm B} =0}$ and ${{\bm K} = 0}$, the two states $\Gamma_5$ and $\Gamma_6$ are guaranteed to be degenerate by time reversal symmetry~\cite{Dresselhaus1955,Herring1937,Elliot1954}. It follows that the levels of excitons bound to a SF are split into sublevels of symmetry ${\Gamma^c_4 \times \Gamma^v_4}$ and ${\Gamma^c_4 \times (\Gamma^v_5 \oplus \Gamma^v_6)}$ where the superscripts $c,v$ refer to the conduction and valence bands. For convenience, the former and latter excitonic states are called the light- and heavy-hole excitons, or the LH and HH excitons, irrespective of the relation between the exciton binding energy and the splitting $\Delta_{HL}$ between the HH and LH sublevels.

The experimental data imply that the in-plane hole $g$-factor in the ground exciton state is negligible. This is exactly true for the HH exciton $\Gamma^c_4 \times (\Gamma^v_5 \oplus \Gamma^v_6)$ at zero magnetic field $B_y$ and zero wavevector ${\bm K}_{\parallel}$. This implies that the splitting $\Delta_{HL}$ is high enough to prevent the mixing between the LH and HH sublevels induced by the finite value of $K_x$ realized in the experiment and by the applied magnetic field, ${|B| \leq 7~\text{T}}$.

In order to confirm that the SF splits HH and LH states, we we can estimate whether the HH-LH splitting is much greater than interactions which mix HH and LH.
The magnetic-field-induced HH-LH mixing can be estimated by using the hole Zeeman Hamiltonian
\begin{equation}
\label{eq:HZ}
{\cal H}_{\rm Z} = - 2\mu_B {\varkappa} \, (\bm J \cdot \bm B),
\end{equation}
where $\mu_B$ is the Bohr magneton, {$\varkappa$ }is the magnetic Luttinger parameter on the order of unity, ${{\bm J} = (J_x,J_y,J_z)}$ are the matrices of the angular momentum 3/2 operator, and we neglect a weak cubic anisotropy~\cite{Sallen2011}. According to Eq.~\eqref{eq:HZ}, the magnitude of the Zeeman HH-LH coupling matrix element at the maximum field of 7~T is $<1$~meV. As for the HH-LH mixing 
caused by the nonzero value of $K_x$, we note that the maximum exciton wavevector measured in our system is 5~$\mu$m$^{-1}$. In this case the terms of the Luttinger-Kohn Hamiltonian that couple the HH and LH excitons have magnitudes less than 2~\(\mu\)eV~\cite{BirPikus}. Thus, we conclude that no significant mixing occurs if the splitting $\Delta_{HL}$ is greater than a few meV. The lack of significant HH-LH mixing in our experiments is in agreement with the HH-LH splitting of 16 meV estimated for interface excitons in polytypic zinc-blende/wurtzite GaAs nanowires~\cite{Spirkoska2012}.

\begin{table*}[htbp]
	 \caption{Character table for the double group of \(C_{3v}\). Double group characters given in Ref.~\cite{Burns}.}
      \begin{tabular}{ >{$}c<{$} | >{$}c<{$}  >{$}c<{$}  >{$}c<{$}  >{$}c<{$}  >{$}c<{$}  >{$}c<{$}  | >{$}l<{$} }
	C_{3v}  & E & \overline E & \begin{aligned} C_3\\ \overline{C}^2_3 \end{aligned}   & \begin{aligned} C_3^2\\ \overline{C}_3 \end{aligned}  & 3 \sigma_v & 3 \overline{\sigma}_v   & \text{Basis Functions} \T\B \\
	\hline
	\Gamma_1 & 1& 1 & 1 & 1 & 1 & 1 & \text{Scalar; z component of vector} \semi z \semi x^2 +y^2 \semi B_x^2 + B_y^2 \semi B_y^3-3B_yB_x^2 \semi I \semi \sigma^{hh}_y    \\
	\Gamma_2 & 1& 1 & 1 & 1 & -1 & -1 & \text{z component of pseudovector} \semi B_z \semi  B_x^3-3B_xB_y^2 \semi \sigma_z^e \semi \sigma_x^{hh}   \semi \sigma_z^{hh}  \semi \sigma_z^{lh} \\
	\Gamma_3 & 2& 2 & -1 & -1 & 0 & 0 & \text{in-plane vector} \semi (K_x,K_y) \semi (-B_y, B_x) \semi (-\sigma_{y}^e,\sigma_{x}^e)  \semi (-\sigma_{y}^{lh},\sigma_{x}^{lh}) \semi  (B_y^2-B_x^2,2B_xB_y) \\
	\Gamma_4 & 2 & -2 & 1 & -1 & 0 & 0 & \text{spin }\frac{1}{2}:  \left( \ket{\tfrac{1}{2},-\tfrac{1}{2}} ,\, \ket{\tfrac{1}{2},\tfrac{1}{2}} \right)  \T\B \\
	\Gamma_5 & 1 & -1 & -1 & 1 & i & -i & \text{Heavy-hole spin `up'} \semi \tfrac{1}{\sqrt 2} \left( \ket{\tfrac{3}{2}, -\tfrac{3}{2} } + i \ket{\tfrac{3}{2}, \tfrac{3}{2} } \right)  \T\B \\
	\Gamma_6 & 1 & -1 & -1 & 1 & -i & i &  \text{Heavy-hole spin `down'}  \semi \tfrac{1}{\sqrt 2} \left( \ket{\tfrac{3}{2}, -\tfrac{3}{2} } - i \ket{\tfrac{3}{2}, \tfrac{3}{2} } \right)   \T\B \\
	\end{tabular}
   \label{tab:chars}
\end{table*}

In order to determine the dipole moment of the SF exciton using Eq.~\ref{eq:props}, it is necessary to estimate the in-plane exciton effective mass, the sum of the electron and hole in-plane effective masses. While the electron mass is isotropic in GaAs, the effective in-plane hole mass depends on the detailed nature of the stacking fault potential.
We note here that HH-LH mixing can affect the in-plane HH effective mass~\cite{Ivchenko1997}. Using second-order perturbation theory, we obtain:
\begin{equation}\label{eq:heavylightcoupling}
\frac{m_o}{m_{hh,||}} = \gamma_1 + \gamma_3 + (4 \gamma_2^2 + 2 \gamma_3^2 ) \frac{\hbar^2}{m_o} \sum_n \frac{ | \bra{LH,n} \hat{k}_z \ket{HH} |^2 }{ E_{HH} - E_{LH,n} } ,
\end{equation}
where the summation goes over all the light-hole states \(n\) (both bound and continuum), $\ket{HH}$ denotes the ground subband HH envelope function along the $z$ axis and \( \ket{LH,n}\) denote the LH envelopes and \({\hat{k}_z = -\partial/\partial z}\) and $\gamma_1,\gamma_2,\gamma_3$ are the Luttinger parameters~\cite{BirPikus}. The sum in Eq.~\ref{eq:heavylightcoupling} is sensitive to the details of the HH and LH envelope functions as well as to the energy positions of the size-quantized levels~\cite{Ivchenko1997,Ikonic1992}.

The estimate of the exciton dipole moment involves the exciton's effective mass, \({M = m_h^* + m_e^*}\), see Eq.~\eqref{eq:props}, where \(m_h^*\) and \(m_e^*\) are the in-plane effective masses of the hole and electron. The electron effective mass is isotropic and therefore its in-plane value is ${m_e^* = 0.067\,m_o}$. Due to the anisotropy of the hole effective mass and the unknown extent of HH-LH coupling, determining the correct value of \(m_h^*\) requires a complex calculation, which we do not perform here. However, we can estimate the HH effective mass using some simple arguments. If HH-LH coupling is neglected, the in-plane heavy hole mass is that of a heavy-hole in the (111) plane, ${m_h^* = 0.10 \,m_o}$~\cite{Ikonic1992}. Another estimate of the in-plane heavy hole mass can be made from the spatially averaged heavy-hole mass of $0.45 \, m_e$. We will use the lower value to estimate the dipole moment, and the higher value to estimate the error in our measurement of the dipole moment. This conservative procedure yields the minimum possible value of the dipole moment. 


%
%

\section{Hamiltonian by symmetry}\label{sec:hamiltonianBySymmetry}

The form of the Hamiltonian describing the SF-bound exciton state can be derived based on the symmetry of the SF system. Specifically, we would like to find terms of the Hamiltonian that are odd with magnetic field and can thus explain the magnetic non-reciprocity data.

It is first necessary to find how the symmetry operations affect a vector, such as a position vector \({\bm r =(x,y,z)}\), and a pseudo-vector, such as magnetic field \({\bm B= (B_x, B_y, B_z)}\). In free space, $\bm r$ and $\bm B$ transform according to the $D^{-}_1$ and $D^{+}_1$ irreducible representations, where $\pm$ denotes parity with respect to space inversion.
Making use of the compatibility tables for \Ctv point symmetry, one can readily check that $D^{-}_1 = \Gamma_1 \oplus \Gamma_3$, while $D^{+}_1 = \Gamma_2\oplus \Gamma_3$. Taking into account that under rotations the components of polar and axial vectors transform identically and making use of Fig.~\ref{fig:c3vOps} we find that $z$ and $B_z$ transform according to the irreducible representations $\Gamma_1$ and $\Gamma_2$, respectively, while the pairs $(x,y)$ and $(-B_y,B_x)$ form equivalent bases of the two-dimensional $\Gamma_3$ irreducible representation of \Ctv, see Table~\ref{tab:chars}.

To apply the method of invariants, we need to establish transformation rules for the basic ${2\times 2}$ matrices acting in the spin subspaces of electrons and heavy-holes. We introduce basic electron matrices $I^e$ (the ${2\times 2}$ unit matrix) and ${\bm \sigma^e=(\sigma_x^e,\sigma_y^e,\sigma_z^e)}$ (Pauli matrices) acting in the basis of $\ket{s=\pm 1/2}$ electron spinors. The decomposition ${\Gamma_4 \otimes \Gamma_4^*= \Gamma_1 + \Gamma_2 + \Gamma_3}$ indicates the ways in which the basic electron spin matrices transform. By calculating the effect of the \Ctv symmetry operators on the matrices, one finds that $I^e$ is invariant, $\sigma_z^e$ transforms according to $\Gamma_2$ and  ${(\sigma_x^e,\sigma_y^e)}$ form a basis of the two-dimensional irreducible representation $\Gamma_3$ with $\sigma_x^e$ and $\sigma_y^e$ transforming equivalently to $B_x$ and $B_y$ respectively (Tab.~\ref{tab:chars}). The basic matrices for the light-hole doublet transform in exactly the same way.

By contrast, the heavy-hole spin doublet transforms according to the reducible representation ${\Gamma_5\oplus \Gamma_6}$. The direct product
\begin{equation}
\label{2reprs}
 (\Gamma_5 \oplus \Gamma_6) \otimes (\Gamma_5 \oplus \Gamma_6)^* = 2 \Gamma_1 \oplus 2 \Gamma_2.
 \end{equation}
indicates that among the four basic matrices, $I^{hh}$, ${\bm \sigma^{hh}=(\sigma_x^{hh},\sigma_y^{hh},\sigma_z^{hh})}$ acting in the space ${|\pm 3/2\rangle}$, two are invariant and two transform as $B_z$~\cite{Sallen2011}. Taking into account that at the mirror reflection $\sigma_v||(\bar 110)$ the matrix $\sigma_y^{hh}$ does not change sign, we find that $I^{hh}$ and $\sigma_y^{hh}$ transform according to $\Gamma_1$ (note that $\sigma_y^{hh}$ changes its sign under time reversal whereas $I^{hh}$ does not), while $\sigma_z^{hh}$, $\sigma_x^{hh}$ transform according to $\Gamma_2$, see Tab.~\ref{tab:chars} and Ref.~\cite{Sallen2011}.

Using the transformation properties of the relevant basis functions (Tab.~\ref{tab:chars}), we can build the effective Hamiltonian.
Any valid term of the Hamiltonian must transform as the identity representation \G{1} and be even under time reversal. In order to know which combinations of basis functions transform as \G{1}, we use the product rules for irreducible representations. 
%
%
From this information, we can build the linear in \(\bm B\) Hamiltonian. For the electron, this analysis produces the result that the in-plane and out-of-plane $g$-factors are potentially different:
\begin{equation}
H_\text{Z,e} = \frac{1}{2} g_{zz}^e \mu_B B_z \sigma_z^e +\frac{1}{2} g_{||}^e \mu_B (B_x \sigma_x^e +B_y \sigma_y^e )
\end{equation}
For the heavy holes, the Hamiltonian takes the form~\cite{Sallen2011}:
\begin{equation}\label{eq:zhh}
H_{\text{Z},hh} = \frac{1}{2}\mu_B B_z \left( g_{zz}^{hh} \sigma_z^{hh} + g_{zx}^{hh} \sigma_x^{hh} \right).
\end{equation}
Eqquation~\ref{eq:zhh} implies that the heavy hole has a zero in-plane $g$-factor, and that an out-of-plane component \(B_z\) creates a tilted effective field with a spin precession vector lying in the \({xz}\) plane. We note that the atomic-scale symmetry of the SF makes the two in-plane directions \(x\) and \(y\) inequivalent.

The magneto-Stark Hamiltonian can be derived by symmetry using a similar procedure. The combination of the two \G{3} irreducible representations $(K_x,K_y)$ and $(-B_y,B_x)$ contains an invariant representation, leading to the Hamiltonian
\[ H_\text{magneto-Stark} = \beta'(  K_x B_y - K_y B_x )\]
Furthermore, the Hamiltonian describing the diamagnetic shift is derived from the invariants \(B_z^2\) and \(B_x^2 + B_y^2\), namely,
\[ H_\text{dia} = \beta_1 B_z^2 + \beta_2 (B_x^2 + B_y^2) .\]
For Eq.~\eqref{eq:KB} of the main text, we take into account only an in-plane field effect and set  \(\beta_2 \equiv \beta\).

We note for completeness that, besides \(B\)-linear, \(B\)-quadratic and $KB$-terms, the effective Hamiltonian also includes \(K\) linear terms, \(\sigma_x^e K_y - \sigma_y^e K_x\), which arise from spin-orbit coupling. Our estimates show that these terms are not significant for the relevant wavevectors and do not lead to the non-reciprocal emission spectra to first-order in \(B\).

%
%

\section{Supplemental Angle Resolved Data}\label{sec:suppData}
\begin{figure*}[htb!]
\includegraphics{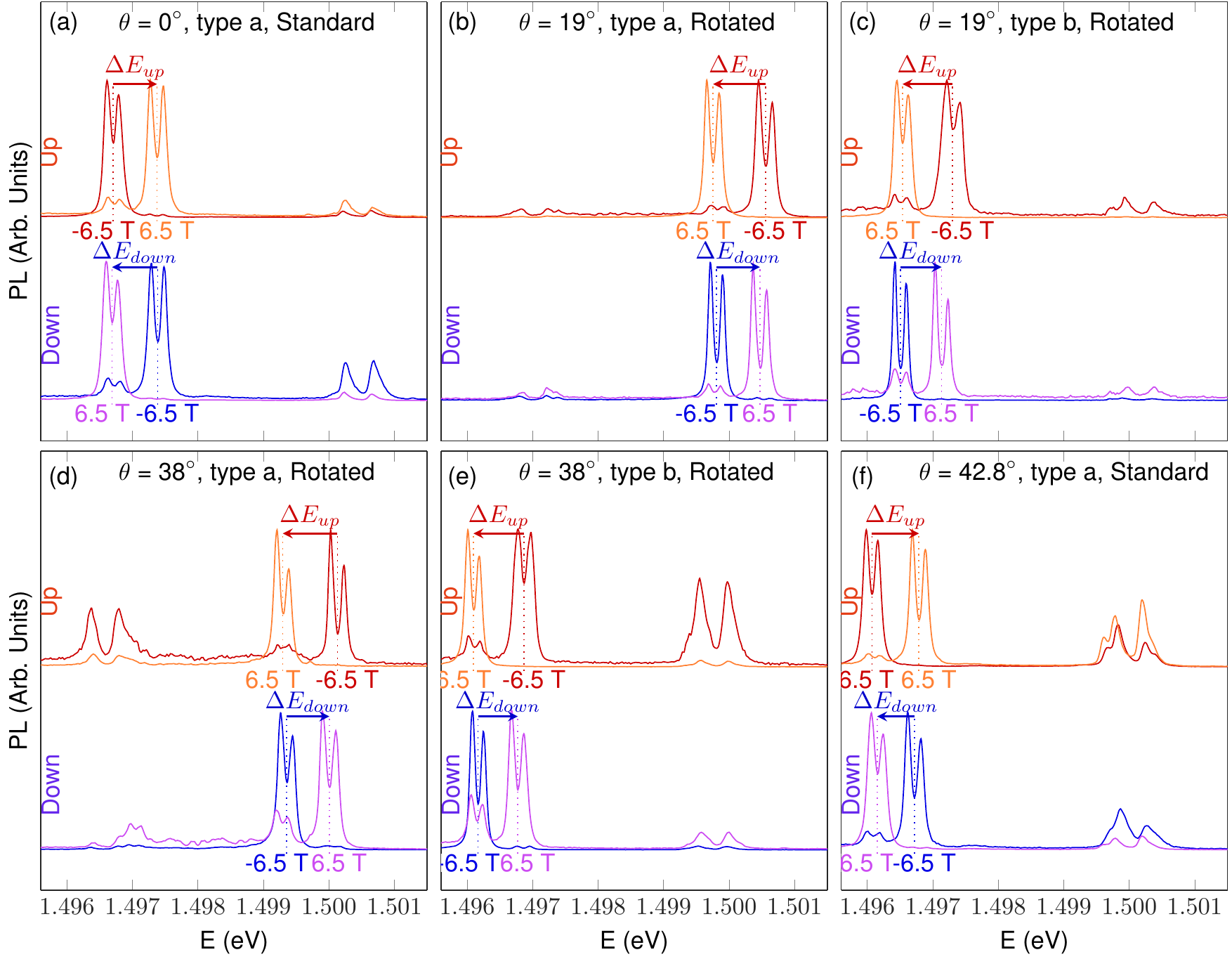}
\caption{Spectra at -6.5 T and 6.5 T for the up and down SF for six different SF pyramid defects. The difference between \(\Delta E_{up}\) and \(\Delta E_{down}\) is visible when \(\theta \neq 0\). 1~$\mu$W, 1.5~K, excite at 1.65~eV, excite and collect H polarization.}
\label{fig:angleAnalysisSpectra}
\end{figure*}

For the angle resolved experiment that tests the magneto-Stark effect, PL spectra were acquired on multiple different SF pyramids positioned at various angles. The spectra of the \emph{up} and \emph{down} SF at \(\pm6.5\)~T are shown in Fig.~\ref{fig:angleAnalysisSpectra}. We note that the PL spectra show a strong and a weak doublet. The origin of the weak doublet is unknown, but we tentatively attribute it to scatter of PL from other exciton populations.

We note that there are two types of SF pyramids: one where the \emph{left/right} SFs show higher energy PL than the \emph{up/down} SFs (type a), and a second type where the emission energies are swapped (type b). For example, using this naming scheme the SF pyramid shown in Fig.~\ref{fig:confocal}(d) is of type a.
Here, the \emph{left/right} directions refer to the direction of the majority of the intrinsic/extrinsic pair defects, visible as single triangles in Fig.~\ref{fig:confocal}(a). We also define a \emph{standard} and \emph{rotated} orientation of the bulk crystal depending on whether the sample is mounted as it is in Fig.~\ref{fig:confocal}(a) (standard) or rotated by \(90^\circ\) about [001] (rotated).

\begin{figure*}[htb!]
\includegraphics{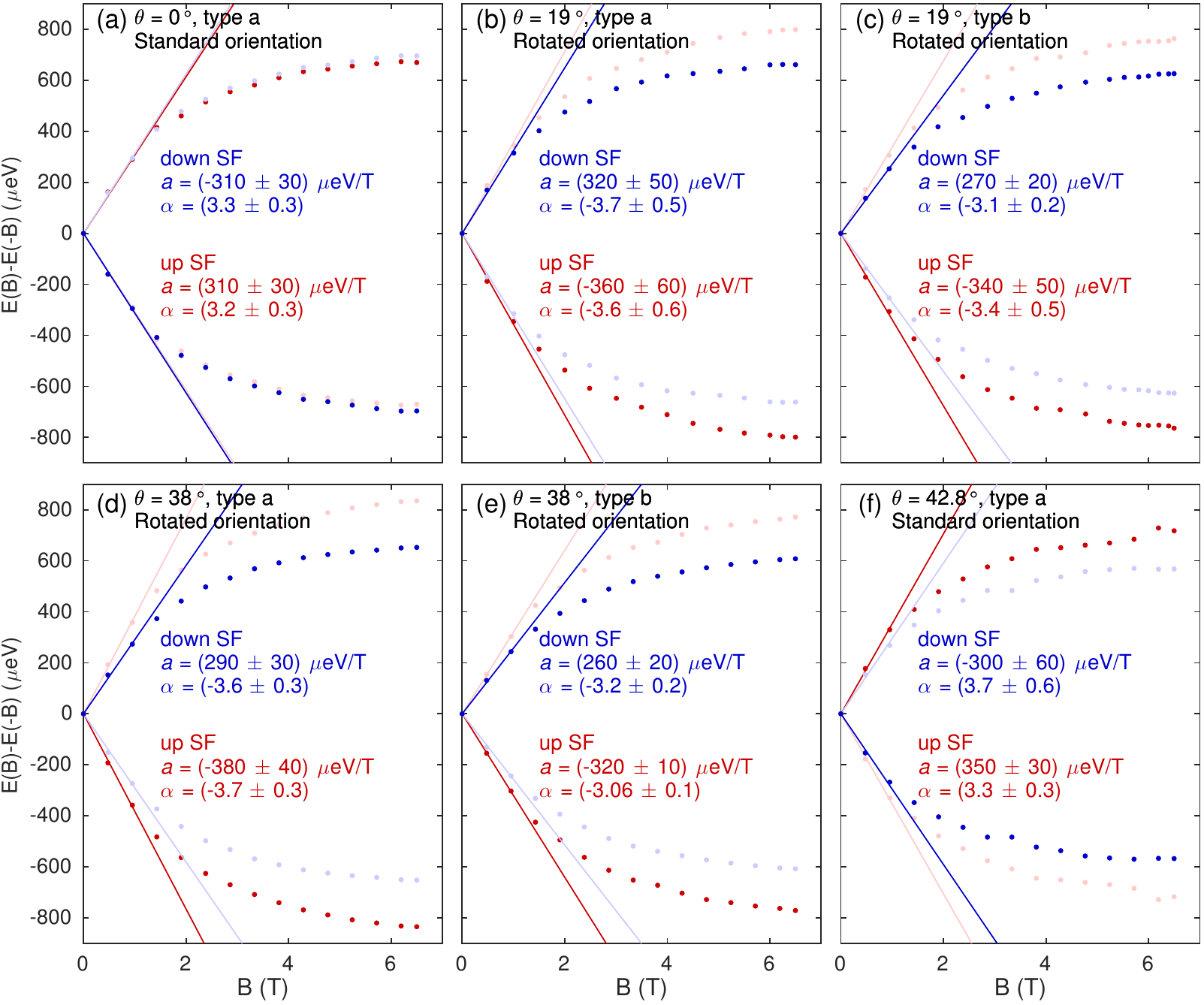}
\caption{The energy of the strong two peaks position in Fig.~\ref{fig:angleAnalysisSpectra} is averaged to make \(E(B)\). Here we plot the difference between the peak position at positive and negative \(B\). The slope at \(B=0\) is extracted with a linear fit (${\Delta E = a B_y}$) to the first three data points. The different panels show the experimental result at a variety of angles. The light colored curves are the negative of the dark colored curves, useful for comparing the absolute value of \(\Delta E\). The mean value of the dimensionless dipole moment in Eq.~\eqref{eq:eupdown} and \eqref{eq:alphan} is \(\alpha = 3.2 \pm 0.2\).}
\label{fig:angleAnalysisDeltaE}
\end{figure*}

The PL emission energy as a function of magnetic field for the 6 SFs is shown in Fig.~\ref{fig:angleAnalysisDeltaE}. These plots show that $d_{he}$, i.e., the electron-hole separation, changes sign for different SF pyramids.   The sign of the splitting \(\Delta E_{up/down}\) reflects the direction of the exciton dipole moment along the SF normal. From Fig.~\ref{fig:angleAnalysisDeltaE}, we find that the atomic structure of the SF pyramid has a reflection symmetry in the $\{110\}$ set of planes, i.e. where the \emph{up} and \emph{down} SFs are interchanged. 


%



\section{Quantitative interpretation of Angle Resolved Data}\label{sec:quantitative}

Combining Eq.~\eqref{eq:anglekagree} and \eqref{eq:magnetostark} of the main text and averaging over two spin states of the exciton's electron, the observed magneto-Stark shift is
\begin{equation}\label{eq:eupdown}
\begin{aligned}
 E_{up} &=  +\alpha \mu_B B_y \sin(\theta_{SF} + \theta'), \\
 E_{down} &= -\alpha \mu_B B_y \sin(\theta_{SF} - \theta')  ,
 \end{aligned}
 \end{equation}
 where the light emission angles are related by Snell's law: \({n \sin\theta' = \sin \theta }\), \({\theta_{SF}=\cos^{-1}\left(1/\sqrt 3 \right) = 54.7^\circ}\) is the angle the SF normal makes with the [001] axis and \({\mu_B = e \hbar/2m_e c}\) is the Bohr magneton (CGS units).
Here \(\alpha\) is a dimensionless measure of the dipole moment:
\begin{equation}\label{eq:alphan}
\alpha = -\frac{en  \hbar \omega}{M c^{2} \mu_B} d_{he} = \frac{-1}{3.2}    \frac{d_{he}}{\text{nm}} ,
\end{equation}
where $d_{he} = \langle z_h-z_e\rangle$  is the hole-electron separation in the exciton, \(z_h\) and \(z_e\) are the hole and electron coordinates and $\langle \ldots \rangle$ denotes the quantum mechanical average (note CGS units). The numerical value uses the estimated in-plane exciton mass described in Sec.~\ref{sec:grouptheory}.
The quantity \({\Delta E_{up/down}(B) = E_{up/down}(B) - E_{up/down}(-B)}\) is the odd part of \(E_{up/down}\) doubled, which is insensitive to overall offsets and the diamagnetic shift.
When \({\theta=\theta'=0}\), PL from the \emph{up} and \emph{down} SFs arises from excitons with the same absolute value of momentum and hence \(\Delta E_{up}\) and \(\Delta E_{down}\) have the same magnitude [Fig.~\ref{fig:angle}(d)]. When the angle is increased, \(\Delta E_{up}\) and \(\Delta E_{down}\)  have different magnitudes [Fig.~\ref{fig:angle}(e)].

In the linear in $B_y$ regime, the magnetic field is not expected to significantly perturb the exciton dipole moment, and so to first order \(\Delta E\) should be proportional to \(B_{y}\). Experimentally, we find the energy shift is linear with \(B_y\) for \({B_y<1~\text{T}}\), and we extract the proportionality constant using a fit to \({\Delta E = aB_y}\). 
The estimate of the dimensionless dipole moment \(\alpha\) depends somewhat on the fitting method used to find the slope \(\Delta E\) vs. \(B_y\) at \({B_y=0}\) from the experimental data. Using different methods, we obtain values of \(\alpha\) ranging from 3.2 to 4.3, and we conservatively use the lowest value of ${\alpha=3.2 \pm 0.3}$ to estimate the exciton dipole moment.
Given the uncertainty in the measurement of \(\alpha\) and the estimation of the in-plane exciton effective mass (Sec.~\ref{sec:grouptheory}), the best guess for the dipole moment is \(10~\text{nm}\), with the true value estimated to fall between \(9\) and \(30~\text{nm}\). 

We note that the combination
\begin{equation}\label{eq:r:s}
r(\theta) = \frac{ |\Delta E_{up}| - |\Delta E_{down}| }{ \frac{1}{2} \big( |\Delta E_{up}| + \left|\Delta E_{down} \right| \big) }
\end{equation}
depends only on given geometry and the index of refraction (to first order in $B$), and is independent of the exciton dipole moment. 
Using Eq.~\ref{eq:eupdown} and expanding about small \(\theta\), we can produce a small-$\theta$ approximation for \(r(\theta)\),
\begin{equation}\label{eq:rapprox}
r(\theta) \approx \frac{ 2\cot{\theta_{SF} }}{n} \theta 
\end{equation}
correct to within 10\% for angles up to 45$^\circ$ and \({n=3.5}\). We obtain good agreement between \(r(\theta)\) calculated experimentally from the \({B=0}\) slope of \(\Delta E\) with Eq.~\ref{eq:r:s}. (Fig.~\ref{fig:angle}h). We emphasize that this model agrees with the data without any fit parameters. %

\bibliography{toddkarin.bib}

\begin{thebibliography}{59}%
\makeatletter
\providecommand \@ifxundefined [1]{%
 \@ifx{#1\undefined}
}%
\providecommand \@ifnum [1]{%
 \ifnum #1\expandafter \@firstoftwo
 \else \expandafter \@secondoftwo
 \fi
}%
\providecommand \@ifx [1]{%
 \ifx #1\expandafter \@firstoftwo
 \else \expandafter \@secondoftwo
 \fi
}%
\providecommand \natexlab [1]{#1}%
\providecommand \enquote  [1]{``#1''}%
\providecommand \bibnamefont  [1]{#1}%
\providecommand \bibfnamefont [1]{#1}%
\providecommand \citenamefont [1]{#1}%
\providecommand \href@noop [0]{\@secondoftwo}%
\providecommand \href [0]{\begingroup \@sanitize@url \@href}%
\providecommand \@href[1]{\@@startlink{#1}\@@href}%
\providecommand \@@href[1]{\endgroup#1\@@endlink}%
\providecommand \@sanitize@url [0]{\catcode `\\12\catcode `\$12\catcode
  `\&12\catcode `\#12\catcode `\^12\catcode `\_12\catcode `\%12\relax}%
\providecommand \@@startlink[1]{}%
\providecommand \@@endlink[0]{}%
\providecommand \url  [0]{\begingroup\@sanitize@url \@url }%
\providecommand \@url [1]{\endgroup\@href {#1}{\urlprefix }}%
\providecommand \urlprefix  [0]{URL }%
\providecommand \Eprint [0]{\href }%
\providecommand \doibase [0]{http://dx.doi.org/}%
\providecommand \selectlanguage [0]{\@gobble}%
\providecommand \bibinfo  [0]{\@secondoftwo}%
\providecommand \bibfield  [0]{\@secondoftwo}%
\providecommand \translation [1]{[#1]}%
\providecommand \BibitemOpen [0]{}%
\providecommand \bibitemStop [0]{}%
\providecommand \bibitemNoStop [0]{.\EOS\space}%
\providecommand \EOS [0]{\spacefactor3000\relax}%
\providecommand \BibitemShut  [1]{\csname bibitem#1\endcsname}%
\let\auto@bib@innerbib\@empty
\bibitem [{\citenamefont {Guha}\ \emph {et~al.}(1993)\citenamefont {Guha},
  \citenamefont {DePuydt}, \citenamefont {Haase}, \citenamefont {Qiu},\ and\
  \citenamefont {Cheng}}]{Guha1993}%
  \BibitemOpen
  \bibfield  {author} {\bibinfo {author} {\bibfnamefont {S.}~\bibnamefont
  {Guha}}, \bibinfo {author} {\bibfnamefont {J.~M.}\ \bibnamefont {DePuydt}},
  \bibinfo {author} {\bibfnamefont {M.~A.}\ \bibnamefont {Haase}}, \bibinfo
  {author} {\bibfnamefont {J.}~\bibnamefont {Qiu}}, \ and\ \bibinfo {author}
  {\bibfnamefont {H.}~\bibnamefont {Cheng}},\ }\bibfield  {title} {\enquote
  {\bibinfo {title} {{Degradation of II-VI based blue-green light emitters}},}\
  }\href {\doibase 10.1063/1.110218} {\bibfield  {journal} {\bibinfo  {journal}
  {Applied Physics Letters}\ }\textbf {\bibinfo {volume} {63}},\ \bibinfo
  {pages} {3107--3109} (\bibinfo {year} {1993})}\BibitemShut {NoStop}%
\bibitem [{\citenamefont {Colli}\ \emph {et~al.}(2003)\citenamefont {Colli},
  \citenamefont {Pelucchi},\ and\ \citenamefont {Franciosi}}]{Colli2003}%
  \BibitemOpen
  \bibfield  {author} {\bibinfo {author} {\bibfnamefont {A.}~\bibnamefont
  {Colli}}, \bibinfo {author} {\bibfnamefont {E.}~\bibnamefont {Pelucchi}}, \
  and\ \bibinfo {author} {\bibfnamefont {A.}~\bibnamefont {Franciosi}},\
  }\bibfield  {title} {\enquote {\bibinfo {title} {{Controlling the native
  stacking fault density in II-VI/III-V heterostructures}},}\ }\href {\doibase
  10.1063/1.1589195} {\bibfield  {journal} {\bibinfo  {journal} {Applied
  Physics Letters}\ }\textbf {\bibinfo {volume} {83}},\ \bibinfo {pages}
  {81--83} (\bibinfo {year} {2003})}\BibitemShut {NoStop}%
\bibitem [{\citenamefont {Caroff}\ \emph {et~al.}(2011)\citenamefont {Caroff},
  \citenamefont {Bolinsson},\ and\ \citenamefont {Johansson}}]{Caroff2011}%
  \BibitemOpen
  \bibfield  {author} {\bibinfo {author} {\bibfnamefont {P.}~\bibnamefont
  {Caroff}}, \bibinfo {author} {\bibfnamefont {J.}~\bibnamefont {Bolinsson}}, \
  and\ \bibinfo {author} {\bibfnamefont {J.}~\bibnamefont {Johansson}},\
  }\bibfield  {title} {\enquote {\bibinfo {title} {{Crystal phases in III--V
  nanowires: from random toward engineered polytypism}},}\ }\href {\doibase
  10.1109/jstqe.2010.2070790} {\bibfield  {journal} {\bibinfo  {journal} {IEEE
  Journal of Selected Topics in Quantum Electronics}\ }\textbf {\bibinfo
  {volume} {17}},\ \bibinfo {pages} {829--846} (\bibinfo {year}
  {2011})}\BibitemShut {NoStop}%
\bibitem [{\citenamefont {Akopian}\ \emph {et~al.}(2010)\citenamefont
  {Akopian}, \citenamefont {Patriarche}, \citenamefont {Liu}, \citenamefont
  {Harmand},\ and\ \citenamefont {Zwiller}}]{Akopian2010}%
  \BibitemOpen
  \bibfield  {author} {\bibinfo {author} {\bibfnamefont {N.}~\bibnamefont
  {Akopian}}, \bibinfo {author} {\bibfnamefont {G.}~\bibnamefont {Patriarche}},
  \bibinfo {author} {\bibfnamefont {L.}~\bibnamefont {Liu}}, \bibinfo {author}
  {\bibfnamefont {J.~C.}\ \bibnamefont {Harmand}}, \ and\ \bibinfo {author}
  {\bibfnamefont {V.}~\bibnamefont {Zwiller}},\ }\bibfield  {title} {\enquote
  {\bibinfo {title} {{Crystal phase quantum dots}},}\ }\href {\doibase
  10.1021/nl903534n} {\bibfield  {journal} {\bibinfo  {journal} {Nano Lett.}\
  }\textbf {\bibinfo {volume} {10}},\ \bibinfo {pages} {1198--1201} (\bibinfo
  {year} {2010})}\BibitemShut {NoStop}%
\bibitem [{\citenamefont {Assali}\ \emph {et~al.}(2013)\citenamefont {Assali},
  \citenamefont {Zardo}, \citenamefont {Plissard}, \citenamefont {Kriegner},
  \citenamefont {Verheijen}, \citenamefont {Bauer}, \citenamefont {Meijerink},
  \citenamefont {Belabbes}, \citenamefont {Bechstedt}, \citenamefont
  {Haverkort},\ and\ \citenamefont {Bakkers}}]{Assali2013}%
  \BibitemOpen
  \bibfield  {author} {\bibinfo {author} {\bibfnamefont {S.}~\bibnamefont
  {Assali}}, \bibinfo {author} {\bibfnamefont {I.}~\bibnamefont {Zardo}},
  \bibinfo {author} {\bibfnamefont {S.}~\bibnamefont {Plissard}}, \bibinfo
  {author} {\bibfnamefont {D.}~\bibnamefont {Kriegner}}, \bibinfo {author}
  {\bibfnamefont {M.~A.}\ \bibnamefont {Verheijen}}, \bibinfo {author}
  {\bibfnamefont {G.}~\bibnamefont {Bauer}}, \bibinfo {author} {\bibfnamefont
  {A.}~\bibnamefont {Meijerink}}, \bibinfo {author} {\bibfnamefont
  {A.}~\bibnamefont {Belabbes}}, \bibinfo {author} {\bibfnamefont
  {F.}~\bibnamefont {Bechstedt}}, \bibinfo {author} {\bibfnamefont {J.~E.~M.}\
  \bibnamefont {Haverkort}}, \ and\ \bibinfo {author} {\bibfnamefont {E.~P.
  A.~M.}\ \bibnamefont {Bakkers}},\ }\bibfield  {title} {\enquote {\bibinfo
  {title} {{Direct band gap wurtzite gallium phosphide nanowires}},}\
  }\bibfield  {booktitle} {\emph {\bibinfo {booktitle} {Nano Letters}},\ }\href
  {\doibase 10.1021/nl304723c} {\bibfield  {journal} {\bibinfo  {journal} {Nano
  Lett.}\ }\textbf {\bibinfo {volume} {13}},\ \bibinfo {pages} {1559--1563}
  (\bibinfo {year} {2013})}\BibitemShut {NoStop}%
\bibitem [{\citenamefont {Butov}\ \emph
  {et~al.}(2002{\natexlab{a}})\citenamefont {Butov}, \citenamefont {Gossard},\
  and\ \citenamefont {Chemla}}]{Butov2002}%
  \BibitemOpen
  \bibfield  {author} {\bibinfo {author} {\bibfnamefont {L.~V.}\ \bibnamefont
  {Butov}}, \bibinfo {author} {\bibfnamefont {A.~C.}\ \bibnamefont {Gossard}},
  \ and\ \bibinfo {author} {\bibfnamefont {D.~S.}\ \bibnamefont {Chemla}},\
  }\bibfield  {title} {\enquote {\bibinfo {title} {{Macroscopically ordered
  state in an exciton system}},}\ }\href {\doibase 10.1038/nature00943}
  {\bibfield  {journal} {\bibinfo  {journal} {Nature}\ }\textbf {\bibinfo
  {volume} {418}},\ \bibinfo {pages} {751--754} (\bibinfo {year}
  {2002}{\natexlab{a}})}\BibitemShut {NoStop}%
\bibitem [{\citenamefont {High}\ \emph {et~al.}(2009)\citenamefont {High},
  \citenamefont {Thomas}, \citenamefont {Grosso}, \citenamefont {Remeika},
  \citenamefont {Hammack}, \citenamefont {Meyertholen}, \citenamefont {Fogler},
  \citenamefont {Butov}, \citenamefont {Hanson},\ and\ \citenamefont
  {Gossard}}]{High2009}%
  \BibitemOpen
  \bibfield  {author} {\bibinfo {author} {\bibfnamefont {A.~A.}\ \bibnamefont
  {High}}, \bibinfo {author} {\bibfnamefont {A.~K.}\ \bibnamefont {Thomas}},
  \bibinfo {author} {\bibfnamefont {G.}~\bibnamefont {Grosso}}, \bibinfo
  {author} {\bibfnamefont {M.}~\bibnamefont {Remeika}}, \bibinfo {author}
  {\bibfnamefont {A.~T.}\ \bibnamefont {Hammack}}, \bibinfo {author}
  {\bibfnamefont {A.~D.}\ \bibnamefont {Meyertholen}}, \bibinfo {author}
  {\bibfnamefont {M.~M.}\ \bibnamefont {Fogler}}, \bibinfo {author}
  {\bibfnamefont {L.~V.}\ \bibnamefont {Butov}}, \bibinfo {author}
  {\bibfnamefont {M.}~\bibnamefont {Hanson}}, \ and\ \bibinfo {author}
  {\bibfnamefont {A.~C.}\ \bibnamefont {Gossard}},\ }\bibfield  {title}
  {\enquote {\bibinfo {title} {{Trapping indirect excitons in a GaAs
  quantum-well structure with a diamond-shaped electrostatic trap}},}\ }\href
  {\doibase 10.1103/PhysRevLett.103.087403} {\bibfield  {journal} {\bibinfo
  {journal} {Phys. Rev. Lett.}\ }\textbf {\bibinfo {volume} {103}},\ \bibinfo
  {pages} {087403} (\bibinfo {year} {2009})}\BibitemShut {NoStop}%
\bibitem [{\citenamefont {Snoke}\ \emph {et~al.}(2002)\citenamefont {Snoke},
  \citenamefont {Denev}, \citenamefont {Liu}, \citenamefont {Pfeiffer},\ and\
  \citenamefont {West}}]{Snoke2002}%
  \BibitemOpen
  \bibfield  {author} {\bibinfo {author} {\bibfnamefont {D.}~\bibnamefont
  {Snoke}}, \bibinfo {author} {\bibfnamefont {S.}~\bibnamefont {Denev}},
  \bibinfo {author} {\bibfnamefont {Y.}~\bibnamefont {Liu}}, \bibinfo {author}
  {\bibfnamefont {L.}~\bibnamefont {Pfeiffer}}, \ and\ \bibinfo {author}
  {\bibfnamefont {K.}~\bibnamefont {West}},\ }\bibfield  {title} {\enquote
  {\bibinfo {title} {{Long-range transport in excitonic dark states in coupled
  quantum wells}},}\ }\href {\doibase 10.1038/nature00940} {\bibfield
  {journal} {\bibinfo  {journal} {Nature}\ }\textbf {\bibinfo {volume} {418}},\
  \bibinfo {pages} {754--757} (\bibinfo {year} {2002})}\BibitemShut {NoStop}%
\bibitem [{\citenamefont {Butov}\ \emph
  {et~al.}(2002{\natexlab{b}})\citenamefont {Butov}, \citenamefont {Lai},
  \citenamefont {Ivanov}, \citenamefont {Gossard},\ and\ \citenamefont
  {Chemla}}]{Butov2002bec}%
  \BibitemOpen
  \bibfield  {author} {\bibinfo {author} {\bibfnamefont {L.~V.}\ \bibnamefont
  {Butov}}, \bibinfo {author} {\bibfnamefont {C.~W.}\ \bibnamefont {Lai}},
  \bibinfo {author} {\bibfnamefont {A.~L.}\ \bibnamefont {Ivanov}}, \bibinfo
  {author} {\bibfnamefont {A.~C.}\ \bibnamefont {Gossard}}, \ and\ \bibinfo
  {author} {\bibfnamefont {D.~S.}\ \bibnamefont {Chemla}},\ }\bibfield  {title}
  {\enquote {\bibinfo {title} {{Towards Bose--Einstein condensation of excitons
  in potential traps}},}\ }\href@noop {} {\bibfield  {journal} {\bibinfo
  {journal} {Nature}\ }\textbf {\bibinfo {volume} {417}},\ \bibinfo {pages}
  {47} (\bibinfo {year} {2002}{\natexlab{b}})}\BibitemShut {NoStop}%
\bibitem [{\citenamefont {Shilo}\ \emph {et~al.}(2013)\citenamefont {Shilo},
  \citenamefont {Cohen}, \citenamefont {Laikhtman}, \citenamefont {West},
  \citenamefont {Pfeiffer},\ and\ \citenamefont {Rapaport}}]{Shilo2013}%
  \BibitemOpen
  \bibfield  {author} {\bibinfo {author} {\bibfnamefont {Yehiel}\ \bibnamefont
  {Shilo}}, \bibinfo {author} {\bibfnamefont {Kobi}\ \bibnamefont {Cohen}},
  \bibinfo {author} {\bibfnamefont {Boris}\ \bibnamefont {Laikhtman}}, \bibinfo
  {author} {\bibfnamefont {Ken}\ \bibnamefont {West}}, \bibinfo {author}
  {\bibfnamefont {Loren}\ \bibnamefont {Pfeiffer}}, \ and\ \bibinfo {author}
  {\bibfnamefont {Ronen}\ \bibnamefont {Rapaport}},\ }\bibfield  {title}
  {\enquote {\bibinfo {title} {{Particle correlations and evidence for dark
  state condensation in a cold dipolar exciton fluid}},}\ }\href {\doibase
  10.1038/ncomms3335} {\bibfield  {journal} {\bibinfo  {journal} {Nat Commun}\
  }\textbf {\bibinfo {volume} {4}} (\bibinfo {year} {2013}),\
  10.1038/ncomms3335}\BibitemShut {NoStop}%
\bibitem [{\citenamefont {High}\ \emph {et~al.}(2013)\citenamefont {High},
  \citenamefont {Hammack}, \citenamefont {Leonard}, \citenamefont {Yang},
  \citenamefont {Butov}, \citenamefont {Ostatnick\'{y}}, \citenamefont
  {Vladimirova}, \citenamefont {Kavokin}, \citenamefont {Liew}, \citenamefont
  {Campman},\ and\ \citenamefont {Gossard}}]{High2013}%
  \BibitemOpen
  \bibfield  {author} {\bibinfo {author} {\bibfnamefont {A.~A.}\ \bibnamefont
  {High}}, \bibinfo {author} {\bibfnamefont {A.~T.}\ \bibnamefont {Hammack}},
  \bibinfo {author} {\bibfnamefont {J.~R.}\ \bibnamefont {Leonard}}, \bibinfo
  {author} {\bibfnamefont {Sen}\ \bibnamefont {Yang}}, \bibinfo {author}
  {\bibfnamefont {L.~V.}\ \bibnamefont {Butov}}, \bibinfo {author}
  {\bibfnamefont {T.}~\bibnamefont {Ostatnick\'{y}}}, \bibinfo {author}
  {\bibfnamefont {M.}~\bibnamefont {Vladimirova}}, \bibinfo {author}
  {\bibfnamefont {A.~V.}\ \bibnamefont {Kavokin}}, \bibinfo {author}
  {\bibfnamefont {T.~C.~H.}\ \bibnamefont {Liew}}, \bibinfo {author}
  {\bibfnamefont {K.~L.}\ \bibnamefont {Campman}}, \ and\ \bibinfo {author}
  {\bibfnamefont {A.~C.}\ \bibnamefont {Gossard}},\ }\bibfield  {title}
  {\enquote {\bibinfo {title} {{Spin currents in a coherent exciton gas}},}\
  }\href {\doibase 10.1103/physrevlett.110.246403} {\bibfield  {journal}
  {\bibinfo  {journal} {Physical Review Letters}\ }\textbf {\bibinfo {volume}
  {110}} (\bibinfo {year} {2013}),\ 10.1103/physrevlett.110.246403}\BibitemShut
  {NoStop}%
\bibitem [{\citenamefont {Fogler}\ \emph {et~al.}(2014)\citenamefont {Fogler},
  \citenamefont {Butov},\ and\ \citenamefont {Novoselov}}]{Fogler2014}%
  \BibitemOpen
  \bibfield  {author} {\bibinfo {author} {\bibfnamefont {M.~M.}\ \bibnamefont
  {Fogler}}, \bibinfo {author} {\bibfnamefont {L.~V.}\ \bibnamefont {Butov}}, \
  and\ \bibinfo {author} {\bibfnamefont {K.~S.}\ \bibnamefont {Novoselov}},\
  }\bibfield  {title} {\enquote {\bibinfo {title} {{High-temperature
  superfluidity with indirect excitons in van der Waals heterostructures}},}\
  }\href {\doibase 10.1038/ncomms5555} {\bibfield  {journal} {\bibinfo
  {journal} {Nature Communications}\ }\textbf {\bibinfo {volume} {5}} (\bibinfo
  {year} {2014}),\ 10.1038/ncomms5555}\BibitemShut {NoStop}%
\bibitem [{\citenamefont {Gorbunov}\ and\ \citenamefont
  {Timofeev}(2006)}]{Gorbunov2006}%
  \BibitemOpen
  \bibfield  {author} {\bibinfo {author} {\bibfnamefont {A.~V.}\ \bibnamefont
  {Gorbunov}}\ and\ \bibinfo {author} {\bibfnamefont {V.~B.}\ \bibnamefont
  {Timofeev}},\ }\bibfield  {title} {\enquote {\bibinfo {title} {{Large-scale
  coherence of the bose condensate of spatially indirect excitons}},}\ }\href
  {\doibase 10.1134/S0021364006180111} {\bibfield  {journal} {\bibinfo
  {journal} {JETP Letters}\ }\textbf {\bibinfo {volume} {84}},\ \bibinfo
  {pages} {329} (\bibinfo {year} {2006})}\BibitemShut {NoStop}%
\bibitem [{\citenamefont {Snoke}(2011)}]{Snoke2011}%
  \BibitemOpen
  \bibfield  {author} {\bibinfo {author} {\bibfnamefont {D.~W.}\ \bibnamefont
  {Snoke}},\ }\bibfield  {title} {\enquote {\bibinfo {title} {{Coherence and
  optical emission from bilayer exciton condensates}},}\ }\href {\doibase
  10.1155/2011/938609} {\bibfield  {journal} {\bibinfo  {journal} {Advances in
  Condensed Matter Physics}\ }\textbf {\bibinfo {volume} {2011}},\ \bibinfo
  {pages} {1--7} (\bibinfo {year} {2011})}\BibitemShut {NoStop}%
\bibitem [{\citenamefont {Nelsen}\ \emph {et~al.}(2009)\citenamefont {Nelsen},
  \citenamefont {Balili}, \citenamefont {Snoke}, \citenamefont {Pfeiffer},\
  and\ \citenamefont {West}}]{Nelson2009}%
  \BibitemOpen
  \bibfield  {author} {\bibinfo {author} {\bibfnamefont {B.}~\bibnamefont
  {Nelsen}}, \bibinfo {author} {\bibfnamefont {R.}~\bibnamefont {Balili}},
  \bibinfo {author} {\bibfnamefont {D.~W.}\ \bibnamefont {Snoke}}, \bibinfo
  {author} {\bibfnamefont {L.}~\bibnamefont {Pfeiffer}}, \ and\ \bibinfo
  {author} {\bibfnamefont {K.}~\bibnamefont {West}},\ }\bibfield  {title}
  {\enquote {\bibinfo {title} {{Lasing and polariton condensation: Two distinct
  transitions in GaAs microcavities with stress traps}},}\ }\href {\doibase
  10.1063/1.3140822} {\bibfield  {journal} {\bibinfo  {journal} {Journal of
  Applied Physics}\ }\textbf {\bibinfo {volume} {105}},\ \bibinfo {pages}
  {122414} (\bibinfo {year} {2009})}\BibitemShut {NoStop}%
\bibitem [{\citenamefont {Balili}\ \emph {et~al.}(2009)\citenamefont {Balili},
  \citenamefont {Nelsen}, \citenamefont {Snoke}, \citenamefont {Pfeiffer},\
  and\ \citenamefont {West}}]{Balili2009}%
  \BibitemOpen
  \bibfield  {author} {\bibinfo {author} {\bibfnamefont {R.}~\bibnamefont
  {Balili}}, \bibinfo {author} {\bibfnamefont {B.}~\bibnamefont {Nelsen}},
  \bibinfo {author} {\bibfnamefont {D.~W.}\ \bibnamefont {Snoke}}, \bibinfo
  {author} {\bibfnamefont {L.}~\bibnamefont {Pfeiffer}}, \ and\ \bibinfo
  {author} {\bibfnamefont {K.}~\bibnamefont {West}},\ }\bibfield  {title}
  {\enquote {\bibinfo {title} {{Role of the stress trap in the polariton
  quasiequilibrium condensation in GaAs microcavities}},}\ }\href {\doibase
  10.1103/PhysRevB.79.075319} {\bibfield  {journal} {\bibinfo  {journal} {Phys.
  Rev. B}\ }\textbf {\bibinfo {volume} {79}},\ \bibinfo {pages} {075319}
  (\bibinfo {year} {2009})}\BibitemShut {NoStop}%
\bibitem [{\citenamefont {High}\ \emph {et~al.}(2012)\citenamefont {High},
  \citenamefont {Leonard}, \citenamefont {Hammack}, \citenamefont {Fogler},
  \citenamefont {Butov}, \citenamefont {Kavokin}, \citenamefont {Campman},\
  and\ \citenamefont {Gossard}}]{High2012}%
  \BibitemOpen
  \bibfield  {author} {\bibinfo {author} {\bibfnamefont {A.~A.}\ \bibnamefont
  {High}}, \bibinfo {author} {\bibfnamefont {J.~R.}\ \bibnamefont {Leonard}},
  \bibinfo {author} {\bibfnamefont {A.~T.}\ \bibnamefont {Hammack}}, \bibinfo
  {author} {\bibfnamefont {M.~M.}\ \bibnamefont {Fogler}}, \bibinfo {author}
  {\bibfnamefont {L.~V.}\ \bibnamefont {Butov}}, \bibinfo {author}
  {\bibfnamefont {A.~V.}\ \bibnamefont {Kavokin}}, \bibinfo {author}
  {\bibfnamefont {K.~L.}\ \bibnamefont {Campman}}, \ and\ \bibinfo {author}
  {\bibfnamefont {A.~C.}\ \bibnamefont {Gossard}},\ }\bibfield  {title}
  {\enquote {\bibinfo {title} {{Spontaneous coherence in a cold exciton
  gas}},}\ }\href {\doibase 10.1038/nature10903} {\bibfield  {journal}
  {\bibinfo  {journal} {Nature}\ }\textbf {\bibinfo {volume} {483}},\ \bibinfo
  {pages} {584--588} (\bibinfo {year} {2012})}\BibitemShut {NoStop}%
\bibitem [{\citenamefont {Amo}\ \emph {et~al.}(2010)\citenamefont {Amo},
  \citenamefont {Liew}, \citenamefont {Adrados}, \citenamefont {Houdre},
  \citenamefont {Giacobino}, \citenamefont {Kavokin},\ and\ \citenamefont
  {Bramati}}]{Amo2010}%
  \BibitemOpen
  \bibfield  {author} {\bibinfo {author} {\bibfnamefont {A.}~\bibnamefont
  {Amo}}, \bibinfo {author} {\bibfnamefont {T.~C.~H.}\ \bibnamefont {Liew}},
  \bibinfo {author} {\bibfnamefont {C.}~\bibnamefont {Adrados}}, \bibinfo
  {author} {\bibfnamefont {R.}~\bibnamefont {Houdre}}, \bibinfo {author}
  {\bibfnamefont {E.}~\bibnamefont {Giacobino}}, \bibinfo {author}
  {\bibfnamefont {A.~V.}\ \bibnamefont {Kavokin}}, \ and\ \bibinfo {author}
  {\bibfnamefont {A.}~\bibnamefont {Bramati}},\ }\bibfield  {title} {\enquote
  {\bibinfo {title} {{Exciton-polariton spin switches}},}\ }\href {\doibase
  10.1038/nphoton.2010.79} {\bibfield  {journal} {\bibinfo  {journal} {Nature
  Photonics}\ }\textbf {\bibinfo {volume} {4}},\ \bibinfo {pages} {361--366}
  (\bibinfo {year} {2010})}\BibitemShut {NoStop}%
\bibitem [{\citenamefont {Nguyen}\ \emph {et~al.}(2013)\citenamefont {Nguyen},
  \citenamefont {Vishnevsky}, \citenamefont {Sturm}, \citenamefont {Tanese},
  \citenamefont {Solnyshkov}, \citenamefont {Galopin}, \citenamefont
  {Lema\^{i}tre}, \citenamefont {Sagnes}, \citenamefont {Amo}, \citenamefont
  {Malpuech},\ and\ \citenamefont {Bloch}}]{Nguyen2013}%
  \BibitemOpen
  \bibfield  {author} {\bibinfo {author} {\bibfnamefont {H.~S.}\ \bibnamefont
  {Nguyen}}, \bibinfo {author} {\bibfnamefont {D.}~\bibnamefont {Vishnevsky}},
  \bibinfo {author} {\bibfnamefont {C.}~\bibnamefont {Sturm}}, \bibinfo
  {author} {\bibfnamefont {D.}~\bibnamefont {Tanese}}, \bibinfo {author}
  {\bibfnamefont {D.}~\bibnamefont {Solnyshkov}}, \bibinfo {author}
  {\bibfnamefont {E.}~\bibnamefont {Galopin}}, \bibinfo {author} {\bibfnamefont
  {A.}~\bibnamefont {Lema\^{i}tre}}, \bibinfo {author} {\bibfnamefont
  {I.}~\bibnamefont {Sagnes}}, \bibinfo {author} {\bibfnamefont
  {A.}~\bibnamefont {Amo}}, \bibinfo {author} {\bibfnamefont {G.}~\bibnamefont
  {Malpuech}}, \ and\ \bibinfo {author} {\bibfnamefont {J.}~\bibnamefont
  {Bloch}},\ }\bibfield  {title} {\enquote {\bibinfo {title} {{Realization of a
  double-barrier resonant tunneling diode for cavity polaritons}},}\ }\href
  {\doibase 10.1103/PhysRevLett.110.236601} {\bibfield  {journal} {\bibinfo
  {journal} {Phys. Rev. Lett.}\ }\textbf {\bibinfo {volume} {110}},\ \bibinfo
  {pages} {236601} (\bibinfo {year} {2013})}\BibitemShut {NoStop}%
\bibitem [{\citenamefont {Kammann}\ \emph {et~al.}(2012)\citenamefont
  {Kammann}, \citenamefont {Liew}, \citenamefont {Ohadi}, \citenamefont
  {Cilibrizzi}, \citenamefont {Tsotsis}, \citenamefont {Hatzopoulos},
  \citenamefont {Savvidis}, \citenamefont {Kavokin},\ and\ \citenamefont
  {Lagoudakis}}]{Kammann2012}%
  \BibitemOpen
  \bibfield  {author} {\bibinfo {author} {\bibfnamefont {E.}~\bibnamefont
  {Kammann}}, \bibinfo {author} {\bibfnamefont {T.~C.~H.}\ \bibnamefont
  {Liew}}, \bibinfo {author} {\bibfnamefont {H.}~\bibnamefont {Ohadi}},
  \bibinfo {author} {\bibfnamefont {P.}~\bibnamefont {Cilibrizzi}}, \bibinfo
  {author} {\bibfnamefont {P.}~\bibnamefont {Tsotsis}}, \bibinfo {author}
  {\bibfnamefont {Z.}~\bibnamefont {Hatzopoulos}}, \bibinfo {author}
  {\bibfnamefont {P.~G.}\ \bibnamefont {Savvidis}}, \bibinfo {author}
  {\bibfnamefont {A.~V.}\ \bibnamefont {Kavokin}}, \ and\ \bibinfo {author}
  {\bibfnamefont {P.~G.}\ \bibnamefont {Lagoudakis}},\ }\bibfield  {title}
  {\enquote {\bibinfo {title} {{Nonlinear optical spin Hall effect and
  long-range spin transport in polariton lasers}},}\ }\href {\doibase
  10.1103/PhysRevLett.109.036404} {\bibfield  {journal} {\bibinfo  {journal}
  {Phys. Rev. Lett.}\ }\textbf {\bibinfo {volume} {109}},\ \bibinfo {pages}
  {036404} (\bibinfo {year} {2012})}\BibitemShut {NoStop}%
\bibitem [{sfs()}]{sfsupplement}%
  \BibitemOpen
  \href@noop {} {}\bibinfo {note} {See Supplemental Material at [URL will be
  inserted by publisher] for details on sample growth, an overview of the
  optical spectroscopy system, details of the linewidth measurements, data for
  the angle resolved experiment, derivation of the magneto-Stark Hamiltonian
  and an explanation of the \(g\)-factor analysis.}\BibitemShut {Stop}%
\bibitem [{\citenamefont {Kasai}\ and\ \citenamefont
  {Kawata}(1998)}]{Kasai1998}%
  \BibitemOpen
  \bibfield  {author} {\bibinfo {author} {\bibfnamefont {J.}~\bibnamefont
  {Kasai}}\ and\ \bibinfo {author} {\bibfnamefont {M.}~\bibnamefont {Kawata}},\
  }\bibfield  {title} {\enquote {\bibinfo {title} {{Microphotoluminescence of
  oval defects in a GaAs layer grown by molecular beam epitaxy}},}\ }\href
  {\doibase 10.1063/1.122352} {\bibfield  {journal} {\bibinfo  {journal}
  {Applied Physics Letters}\ }\textbf {\bibinfo {volume} {73}},\ \bibinfo
  {pages} {2012--2014} (\bibinfo {year} {1998})}\BibitemShut {NoStop}%
\bibitem [{\citenamefont {L\"{a}hnemann}\ \emph {et~al.}(2014)\citenamefont
  {L\"{a}hnemann}, \citenamefont {Jahn}, \citenamefont {Brandt}, \citenamefont
  {Flissikowski}, \citenamefont {Dogan},\ and\ \citenamefont
  {Grahn}}]{Lahnemann2014}%
  \BibitemOpen
  \bibfield  {author} {\bibinfo {author} {\bibfnamefont {J.}~\bibnamefont
  {L\"{a}hnemann}}, \bibinfo {author} {\bibfnamefont {U.}~\bibnamefont {Jahn}},
  \bibinfo {author} {\bibfnamefont {O.}~\bibnamefont {Brandt}}, \bibinfo
  {author} {\bibfnamefont {T.}~\bibnamefont {Flissikowski}}, \bibinfo {author}
  {\bibfnamefont {P.}~\bibnamefont {Dogan}}, \ and\ \bibinfo {author}
  {\bibfnamefont {H.~T.}\ \bibnamefont {Grahn}},\ }\bibfield  {title} {\enquote
  {\bibinfo {title} {{Luminescence associated with stacking faults in GaN}},}\
  }\href {\doibase 10.1088/0022-3727/47/42/423001} {\bibfield  {journal}
  {\bibinfo  {journal} {Journal of Physics D: Applied Physics}\ }\textbf
  {\bibinfo {volume} {47}},\ \bibinfo {pages} {423001} (\bibinfo {year}
  {2014})}\BibitemShut {NoStop}%
\bibitem [{\citenamefont {Algra}\ \emph {et~al.}(2008)\citenamefont {Algra},
  \citenamefont {Verheijen}, \citenamefont {Borgstrom}, \citenamefont {Feiner},
  \citenamefont {Immink}, \citenamefont {van Enckevort}, \citenamefont
  {Vlieg},\ and\ \citenamefont {Bakkers}}]{Algra2008}%
  \BibitemOpen
  \bibfield  {author} {\bibinfo {author} {\bibfnamefont {R.~E.}\ \bibnamefont
  {Algra}}, \bibinfo {author} {\bibfnamefont {M.~A.}\ \bibnamefont
  {Verheijen}}, \bibinfo {author} {\bibfnamefont {M.~T.}\ \bibnamefont
  {Borgstrom}}, \bibinfo {author} {\bibfnamefont {L.~F.}\ \bibnamefont
  {Feiner}}, \bibinfo {author} {\bibfnamefont {G.}~\bibnamefont {Immink}},
  \bibinfo {author} {\bibfnamefont {W.~J.~P.}\ \bibnamefont {van Enckevort}},
  \bibinfo {author} {\bibfnamefont {E.}~\bibnamefont {Vlieg}}, \ and\ \bibinfo
  {author} {\bibfnamefont {E.~P. A.~M.}\ \bibnamefont {Bakkers}},\ }\bibfield
  {title} {\enquote {\bibinfo {title} {{Twinning superlattices in indium
  phosphide nanowires}},}\ }\href {\doibase 10.1038/nature07570} {\bibfield
  {journal} {\bibinfo  {journal} {Nature}\ }\textbf {\bibinfo {volume} {456}},\
  \bibinfo {pages} {369--372} (\bibinfo {year} {2008})}\BibitemShut {NoStop}%
\bibitem [{\citenamefont {Belabbes}\ \emph {et~al.}(2012)\citenamefont
  {Belabbes}, \citenamefont {Panse}, \citenamefont {Furthm\"{u}ller},\ and\
  \citenamefont {Bechstedt}}]{Belabbes2012}%
  \BibitemOpen
  \bibfield  {author} {\bibinfo {author} {\bibfnamefont {A.}~\bibnamefont
  {Belabbes}}, \bibinfo {author} {\bibfnamefont {C.}~\bibnamefont {Panse}},
  \bibinfo {author} {\bibfnamefont {J.}~\bibnamefont {Furthm\"{u}ller}}, \ and\
  \bibinfo {author} {\bibfnamefont {F.}~\bibnamefont {Bechstedt}},\ }\bibfield
  {title} {\enquote {\bibinfo {title} {{Electronic bands of III-V semiconductor
  polytypes and their alignment}},}\ }\href {\doibase
  10.1103/PhysRevB.86.075208} {\bibfield  {journal} {\bibinfo  {journal} {Phys.
  Rev. B}\ }\textbf {\bibinfo {volume} {86}},\ \bibinfo {pages} {075208}
  (\bibinfo {year} {2012})}\BibitemShut {NoStop}%
\bibitem [{\citenamefont {Spirkoska}\ \emph {et~al.}(2009)\citenamefont
  {Spirkoska}, \citenamefont {Arbiol}, \citenamefont {Gustafsson},
  \citenamefont {Conesa-Boj}, \citenamefont {Glas}, \citenamefont {Zardo},
  \citenamefont {Heigoldt}, \citenamefont {Gass}, \citenamefont {Bleloch},
  \citenamefont {Estrade}, \citenamefont {Kaniber}, \citenamefont {Rossler},
  \citenamefont {Peiro}, \citenamefont {Morante}, \citenamefont {Abstreiter},
  \citenamefont {Samuelson},\ and\ \citenamefont
  {Fontcuberta}}]{Spirkoska2009}%
  \BibitemOpen
  \bibfield  {author} {\bibinfo {author} {\bibfnamefont {D.}~\bibnamefont
  {Spirkoska}}, \bibinfo {author} {\bibfnamefont {J.}~\bibnamefont {Arbiol}},
  \bibinfo {author} {\bibfnamefont {A.}~\bibnamefont {Gustafsson}}, \bibinfo
  {author} {\bibfnamefont {S.}~\bibnamefont {Conesa-Boj}}, \bibinfo {author}
  {\bibfnamefont {F.}~\bibnamefont {Glas}}, \bibinfo {author} {\bibfnamefont
  {I.}~\bibnamefont {Zardo}}, \bibinfo {author} {\bibfnamefont
  {M.}~\bibnamefont {Heigoldt}}, \bibinfo {author} {\bibfnamefont {M.~H.}\
  \bibnamefont {Gass}}, \bibinfo {author} {\bibfnamefont {A.~L.}\ \bibnamefont
  {Bleloch}}, \bibinfo {author} {\bibfnamefont {S.}~\bibnamefont {Estrade}},
  \bibinfo {author} {\bibfnamefont {M.}~\bibnamefont {Kaniber}}, \bibinfo
  {author} {\bibfnamefont {J.}~\bibnamefont {Rossler}}, \bibinfo {author}
  {\bibfnamefont {F.}~\bibnamefont {Peiro}}, \bibinfo {author} {\bibfnamefont
  {J.~R.}\ \bibnamefont {Morante}}, \bibinfo {author} {\bibfnamefont
  {G.}~\bibnamefont {Abstreiter}}, \bibinfo {author} {\bibfnamefont
  {L.}~\bibnamefont {Samuelson}}, \ and\ \bibinfo {author} {\bibnamefont
  {Fontcuberta}},\ }\bibfield  {title} {\enquote {\bibinfo {title} {{Structural
  and optical properties of high quality zinc-blende/wurtzite GaAs nanowire
  heterostructures}},}\ }\href {\doibase 10.1103/PhysRevB.80.245325} {\bibfield
   {journal} {\bibinfo  {journal} {Phys. Rev. B}\ }\textbf {\bibinfo {volume}
  {80}},\ \bibinfo {pages} {245325} (\bibinfo {year} {2009})}\BibitemShut
  {NoStop}%
\bibitem [{\citenamefont {Heiss}\ \emph {et~al.}(2011)\citenamefont {Heiss},
  \citenamefont {Conesa-Boj}, \citenamefont {Ren}, \citenamefont {Tseng},
  \citenamefont {Gali}, \citenamefont {Rudolph}, \citenamefont {Uccelli},
  \citenamefont {Peir\'{o}}, \citenamefont {Morante}, \citenamefont {Schuh},
  \citenamefont {Reiger}, \citenamefont {Kaxiras}, \citenamefont {Arbiol},\
  and\ \citenamefont {Fontcuberta}}]{Heiss2011}%
  \BibitemOpen
  \bibfield  {author} {\bibinfo {author} {\bibfnamefont {M.}~\bibnamefont
  {Heiss}}, \bibinfo {author} {\bibfnamefont {S.}~\bibnamefont {Conesa-Boj}},
  \bibinfo {author} {\bibfnamefont {J.}~\bibnamefont {Ren}}, \bibinfo {author}
  {\bibfnamefont {H.~H.}\ \bibnamefont {Tseng}}, \bibinfo {author}
  {\bibfnamefont {A.}~\bibnamefont {Gali}}, \bibinfo {author} {\bibfnamefont
  {A.}~\bibnamefont {Rudolph}}, \bibinfo {author} {\bibfnamefont
  {E.}~\bibnamefont {Uccelli}}, \bibinfo {author} {\bibfnamefont
  {F.}~\bibnamefont {Peir\'{o}}}, \bibinfo {author} {\bibfnamefont {J.~R.}\
  \bibnamefont {Morante}}, \bibinfo {author} {\bibfnamefont {D.}~\bibnamefont
  {Schuh}}, \bibinfo {author} {\bibfnamefont {E.}~\bibnamefont {Reiger}},
  \bibinfo {author} {\bibfnamefont {E.}~\bibnamefont {Kaxiras}}, \bibinfo
  {author} {\bibfnamefont {J.}~\bibnamefont {Arbiol}}, \ and\ \bibinfo {author}
  {\bibnamefont {Fontcuberta}},\ }\bibfield  {title} {\enquote {\bibinfo
  {title} {{Direct correlation of crystal structure and optical properties in
  wurtzite/zinc-blende GaAs nanowire heterostructures}},}\ }\href {\doibase
  10.1103/PhysRevB.83.045303} {\bibfield  {journal} {\bibinfo  {journal} {Phys.
  Rev. B}\ }\textbf {\bibinfo {volume} {83}},\ \bibinfo {pages} {045303}
  (\bibinfo {year} {2011})}\BibitemShut {NoStop}%
\bibitem [{\citenamefont {L\"{a}hnemann}\ \emph {et~al.}(2012)\citenamefont
  {L\"{a}hnemann}, \citenamefont {Brandt}, \citenamefont {Jahn}, \citenamefont
  {Pf\"{u}ller}, \citenamefont {Roder}, \citenamefont {Dogan}, \citenamefont
  {Grosse}, \citenamefont {Belabbes}, \citenamefont {Bechstedt}, \citenamefont
  {Trampert},\ and\ \citenamefont {Geelhaar}}]{Lahnemann2012}%
  \BibitemOpen
  \bibfield  {author} {\bibinfo {author} {\bibfnamefont {J.}~\bibnamefont
  {L\"{a}hnemann}}, \bibinfo {author} {\bibfnamefont {O.}~\bibnamefont
  {Brandt}}, \bibinfo {author} {\bibfnamefont {U.}~\bibnamefont {Jahn}},
  \bibinfo {author} {\bibfnamefont {C.}~\bibnamefont {Pf\"{u}ller}}, \bibinfo
  {author} {\bibfnamefont {C.}~\bibnamefont {Roder}}, \bibinfo {author}
  {\bibfnamefont {P.}~\bibnamefont {Dogan}}, \bibinfo {author} {\bibfnamefont
  {F.}~\bibnamefont {Grosse}}, \bibinfo {author} {\bibfnamefont
  {A.}~\bibnamefont {Belabbes}}, \bibinfo {author} {\bibfnamefont
  {F.}~\bibnamefont {Bechstedt}}, \bibinfo {author} {\bibfnamefont
  {A.}~\bibnamefont {Trampert}}, \ and\ \bibinfo {author} {\bibfnamefont
  {L.}~\bibnamefont {Geelhaar}},\ }\bibfield  {title} {\enquote {\bibinfo
  {title} {{Direct experimental determination of the spontaneous polarization
  of GaN}},}\ }\href {\doibase 10.1103/PhysRevB.86.081302} {\bibfield
  {journal} {\bibinfo  {journal} {Phys. Rev. B}\ }\textbf {\bibinfo {volume}
  {86}},\ \bibinfo {pages} {081302} (\bibinfo {year} {2012})}\BibitemShut
  {NoStop}%
\bibitem [{\citenamefont {Vainorius}\ \emph {et~al.}(2015)\citenamefont
  {Vainorius}, \citenamefont {Lehmann}, \citenamefont {Jacobsson},
  \citenamefont {Samuelson}, \citenamefont {Dick},\ and\ \citenamefont
  {Pistol}}]{Vainorius2015}%
  \BibitemOpen
  \bibfield  {author} {\bibinfo {author} {\bibfnamefont {N.}~\bibnamefont
  {Vainorius}}, \bibinfo {author} {\bibfnamefont {S.}~\bibnamefont {Lehmann}},
  \bibinfo {author} {\bibfnamefont {D.}~\bibnamefont {Jacobsson}}, \bibinfo
  {author} {\bibfnamefont {L.}~\bibnamefont {Samuelson}}, \bibinfo {author}
  {\bibfnamefont {K.~A.}\ \bibnamefont {Dick}}, \ and\ \bibinfo {author}
  {\bibfnamefont {M.~E.}\ \bibnamefont {Pistol}},\ }\bibfield  {title}
  {\enquote {\bibinfo {title} {{Confinement in thickness-controlled GaAs
  polytype nanodots}},}\ }\href {\doibase 10.1021/acs.nanolett.5b00253}
  {\bibfield  {journal} {\bibinfo  {journal} {Nano Lett.}\ }\textbf {\bibinfo
  {volume} {15}},\ \bibinfo {pages} {2652--2656} (\bibinfo {year}
  {2015})}\BibitemShut {NoStop}%
\bibitem [{\citenamefont {Fung}\ \emph {et~al.}(1997)\citenamefont {Fung},
  \citenamefont {Wang},\ and\ \citenamefont {Sou}}]{Fung1997}%
  \BibitemOpen
  \bibfield  {author} {\bibinfo {author} {\bibfnamefont {K.~K.}\ \bibnamefont
  {Fung}}, \bibinfo {author} {\bibfnamefont {N.}~\bibnamefont {Wang}}, \ and\
  \bibinfo {author} {\bibfnamefont {I.~K.}\ \bibnamefont {Sou}},\ }\bibfield
  {title} {\enquote {\bibinfo {title} {{Direct observation of stacking fault
  tetrahedra in ZnSe/GaAs(001) pseudomorphic epilayers by weak beam dark-field
  transmission electron microscopy}},}\ }\href {\doibase 10.1063/1.119858}
  {\bibfield  {journal} {\bibinfo  {journal} {Applied Physics Letters}\
  }\textbf {\bibinfo {volume} {71}},\ \bibinfo {pages} {1225--1227} (\bibinfo
  {year} {1997})}\BibitemShut {NoStop}%
\bibitem [{\citenamefont {Wang}\ \emph {et~al.}(2000)\citenamefont {Wang},
  \citenamefont {Fung},\ and\ \citenamefont {Sou}}]{Wang2000}%
  \BibitemOpen
  \bibfield  {author} {\bibinfo {author} {\bibfnamefont {N.}~\bibnamefont
  {Wang}}, \bibinfo {author} {\bibfnamefont {K.~K.}\ \bibnamefont {Fung}}, \
  and\ \bibinfo {author} {\bibfnamefont {I.~K.}\ \bibnamefont {Sou}},\
  }\bibfield  {title} {\enquote {\bibinfo {title} {{Direct observation of
  stacking fault nucleation in the early stage of ZnSe/GaAs pseudomorphic
  epitaxial layer growth}},}\ }\href {\doibase 10.1063/1.1321732} {\bibfield
  {journal} {\bibinfo  {journal} {Applied Physics Letters}\ }\textbf {\bibinfo
  {volume} {77}},\ \bibinfo {pages} {2846--2848} (\bibinfo {year}
  {2000})}\BibitemShut {NoStop}%
\bibitem [{\citenamefont {Corfdir}\ and\ \citenamefont
  {Lefebvre}(2012)}]{Corfdir2012}%
  \BibitemOpen
  \bibfield  {author} {\bibinfo {author} {\bibfnamefont {P.}~\bibnamefont
  {Corfdir}}\ and\ \bibinfo {author} {\bibfnamefont {P.}~\bibnamefont
  {Lefebvre}},\ }\bibfield  {title} {\enquote {\bibinfo {title} {{Importance of
  excitonic effects and the question of internal electric fields in stacking
  faults and crystal phase quantum discs: The model-case of GaN}},}\ }\href
  {\doibase 10.1063/1.4749789} {\bibfield  {journal} {\bibinfo  {journal}
  {Journal of Applied Physics}\ }\textbf {\bibinfo {volume} {112}},\ \bibinfo
  {pages} {053512} (\bibinfo {year} {2012})}\BibitemShut {NoStop}%
\bibitem [{\citenamefont {Kakibayashi}\ \emph {et~al.}(1984)\citenamefont
  {Kakibayashi}, \citenamefont {Nagata}, \citenamefont {Katayama},\ and\
  \citenamefont {Shiraki}}]{Kakibayashi1984}%
  \BibitemOpen
  \bibfield  {author} {\bibinfo {author} {\bibfnamefont {H.}~\bibnamefont
  {Kakibayashi}}, \bibinfo {author} {\bibfnamefont {F.}~\bibnamefont {Nagata}},
  \bibinfo {author} {\bibfnamefont {Y.}~\bibnamefont {Katayama}}, \ and\
  \bibinfo {author} {\bibfnamefont {Y.}~\bibnamefont {Shiraki}},\ }\bibfield
  {title} {\enquote {\bibinfo {title} {{Structure analysis of oval defect on
  molecular beam epitaxial GaAs layer by cross-sectional transmission electron
  microscopy observation}},}\ }\href {\doibase 10.1143/jjap.23.l846} {\bibfield
   {journal} {\bibinfo  {journal} {Japanese Journal of Applied Physics}\
  }\textbf {\bibinfo {volume} {23}},\ \bibinfo {pages} {L846--L848} (\bibinfo
  {year} {1984})}\BibitemShut {NoStop}%
\bibitem [{\citenamefont {Komatsu}(1988)}]{Komatsu1988}%
  \BibitemOpen
  \bibfield  {author} {\bibinfo {author} {\bibfnamefont {Teruo}\ \bibnamefont
  {Komatsu}},\ }\bibfield  {title} {\enquote {\bibinfo {title} {{Optical
  properties of excitons confined two-dimensionally in a stacking fault plane
  in BiI3}},}\ }\href {\doibase 10.1016/0022-2313(88)90299-2} {\bibfield
  {journal} {\bibinfo  {journal} {Journal of Luminescence}\ }\textbf {\bibinfo
  {volume} {40-41}},\ \bibinfo {pages} {495--496} (\bibinfo {year}
  {1988})}\BibitemShut {NoStop}%
\bibitem [{\citenamefont {Poltavtsev}\ \emph {et~al.}(2014)\citenamefont
  {Poltavtsev}, \citenamefont {Efimov}, \citenamefont {Dolgikh}, \citenamefont
  {Eliseev}, \citenamefont {Petrov},\ and\ \citenamefont
  {Ovsyankin}}]{Poltavtsev2014}%
  \BibitemOpen
  \bibfield  {author} {\bibinfo {author} {\bibfnamefont {S.~V.}\ \bibnamefont
  {Poltavtsev}}, \bibinfo {author} {\bibfnamefont {Yu}~\bibnamefont {Efimov}},
  \bibinfo {author} {\bibfnamefont {Yu}~\bibnamefont {Dolgikh}}, \bibinfo
  {author} {\bibfnamefont {S.~A.}\ \bibnamefont {Eliseev}}, \bibinfo {author}
  {\bibfnamefont {V.~V.}\ \bibnamefont {Petrov}}, \ and\ \bibinfo {author}
  {\bibfnamefont {V.~V.}\ \bibnamefont {Ovsyankin}},\ }\bibfield  {title}
  {\enquote {\bibinfo {title} {{Extremely low inhomogeneous broadening of
  exciton lines in shallow (In,Ga)As/GaAs quantum wells}},}\ }\href {\doibase
  10.1016/j.ssc.2014.09.005} {\bibfield  {journal} {\bibinfo  {journal} {Solid
  State Communications}\ }\textbf {\bibinfo {volume} {199}},\ \bibinfo {pages}
  {47--51} (\bibinfo {year} {2014})}\BibitemShut {NoStop}%
\bibitem [{\citenamefont {Graham}\ \emph {et~al.}(2013)\citenamefont {Graham},
  \citenamefont {Corfdir}, \citenamefont {Heiss}, \citenamefont {Conesa-Boj},
  \citenamefont {Uccelli}, \citenamefont {Fontcuberta},\ and\ \citenamefont
  {Phillips}}]{Graham2013}%
  \BibitemOpen
  \bibfield  {author} {\bibinfo {author} {\bibfnamefont {A.~M.}\ \bibnamefont
  {Graham}}, \bibinfo {author} {\bibfnamefont {P.}~\bibnamefont {Corfdir}},
  \bibinfo {author} {\bibfnamefont {M.}~\bibnamefont {Heiss}}, \bibinfo
  {author} {\bibfnamefont {S.}~\bibnamefont {Conesa-Boj}}, \bibinfo {author}
  {\bibfnamefont {E.}~\bibnamefont {Uccelli}}, \bibinfo {author} {\bibnamefont
  {Fontcuberta}}, \ and\ \bibinfo {author} {\bibfnamefont {R.~T.}\ \bibnamefont
  {Phillips}},\ }\bibfield  {title} {\enquote {\bibinfo {title} {{Exciton
  localization mechanisms in wurtzite/zinc-blende GaAs nanowires}},}\ }\href
  {\doibase 10.1103/PhysRevB.87.125304} {\bibfield  {journal} {\bibinfo
  {journal} {Phys. Rev. B}\ }\textbf {\bibinfo {volume} {87}},\ \bibinfo
  {pages} {125304} (\bibinfo {year} {2013})}\BibitemShut {NoStop}%
\bibitem [{\citenamefont {Pal}\ \emph {et~al.}(2008)\citenamefont {Pal},
  \citenamefont {Goto}, \citenamefont {Ikezawa}, \citenamefont {Masumoto},
  \citenamefont {Mohan}, \citenamefont {Motohisa},\ and\ \citenamefont
  {Fukui}}]{Pal2008}%
  \BibitemOpen
  \bibfield  {author} {\bibinfo {author} {\bibfnamefont {B.}~\bibnamefont
  {Pal}}, \bibinfo {author} {\bibfnamefont {K.}~\bibnamefont {Goto}}, \bibinfo
  {author} {\bibfnamefont {M.}~\bibnamefont {Ikezawa}}, \bibinfo {author}
  {\bibfnamefont {Y.}~\bibnamefont {Masumoto}}, \bibinfo {author}
  {\bibfnamefont {P.}~\bibnamefont {Mohan}}, \bibinfo {author} {\bibfnamefont
  {J.}~\bibnamefont {Motohisa}}, \ and\ \bibinfo {author} {\bibfnamefont
  {T.}~\bibnamefont {Fukui}},\ }\bibfield  {title} {\enquote {\bibinfo {title}
  {{Type-II behavior in wurtzite InP/InAs/InP core-multishell nanowires}},}\
  }\href {\doibase 10.1063/1.2966343} {\bibfield  {journal} {\bibinfo
  {journal} {Applied Physics Letters}\ }\textbf {\bibinfo {volume} {93}},\
  \bibinfo {pages} {073105} (\bibinfo {year} {2008})}\BibitemShut {NoStop}%
\bibitem [{\citenamefont {Signorello}\ \emph {et~al.}(2014)\citenamefont
  {Signorello}, \citenamefont {L\"{o}rtscher}, \citenamefont {Khomyakov},
  \citenamefont {Karg}, \citenamefont {Dheeraj}, \citenamefont {Gotsmann},
  \citenamefont {Weman},\ and\ \citenamefont {Riel}}]{Signorello2014}%
  \BibitemOpen
  \bibfield  {author} {\bibinfo {author} {\bibfnamefont {G.}~\bibnamefont
  {Signorello}}, \bibinfo {author} {\bibfnamefont {E.}~\bibnamefont
  {L\"{o}rtscher}}, \bibinfo {author} {\bibfnamefont {P.~A.}\ \bibnamefont
  {Khomyakov}}, \bibinfo {author} {\bibfnamefont {S.}~\bibnamefont {Karg}},
  \bibinfo {author} {\bibfnamefont {D.~L.}\ \bibnamefont {Dheeraj}}, \bibinfo
  {author} {\bibfnamefont {B.}~\bibnamefont {Gotsmann}}, \bibinfo {author}
  {\bibfnamefont {H.}~\bibnamefont {Weman}}, \ and\ \bibinfo {author}
  {\bibfnamefont {H.}~\bibnamefont {Riel}},\ }\bibfield  {title} {\enquote
  {\bibinfo {title} {{Inducing a direct-to-pseudodirect bandgap transition in
  wurtzite GaAs nanowires with uniaxial stress}},}\ }\href {\doibase
  10.1038/ncomms4655} {\bibfield  {journal} {\bibinfo  {journal} {Nature
  Communications}\ }\textbf {\bibinfo {volume} {5}} (\bibinfo {year} {2014}),\
  10.1038/ncomms4655}\BibitemShut {NoStop}%
\bibitem [{\citenamefont {Spirkoska}\ \emph {et~al.}(2012)\citenamefont
  {Spirkoska}, \citenamefont {Efros}, \citenamefont {Lambrecht}, \citenamefont
  {Cheiwchanchamnangij}, \citenamefont {Morral},\ and\ \citenamefont
  {Abstreiter}}]{Spirkoska2012}%
  \BibitemOpen
  \bibfield  {author} {\bibinfo {author} {\bibfnamefont {D.}~\bibnamefont
  {Spirkoska}}, \bibinfo {author} {\bibfnamefont {A.~L.}\ \bibnamefont
  {Efros}}, \bibinfo {author} {\bibfnamefont {W.~R.~L.}\ \bibnamefont
  {Lambrecht}}, \bibinfo {author} {\bibfnamefont {T.}~\bibnamefont
  {Cheiwchanchamnangij}}, \bibinfo {author} {\bibfnamefont {A.~F.}\
  \bibnamefont {Morral}}, \ and\ \bibinfo {author} {\bibfnamefont
  {G.}~\bibnamefont {Abstreiter}},\ }\bibfield  {title} {\enquote {\bibinfo
  {title} {{Valence band structure of polytypic zinc-blende/wurtzite GaAs
  nanowires probed by polarization-dependent photoluminescence}},}\ }\href
  {\doibase 10.1103/PhysRevB.85.045309} {\bibfield  {journal} {\bibinfo
  {journal} {Phys. Rev. B}\ }\textbf {\bibinfo {volume} {85}},\ \bibinfo
  {pages} {045309} (\bibinfo {year} {2012})}\BibitemShut {NoStop}%
\bibitem [{Note1()}]{Note1}%
  \BibitemOpen
  \bibinfo {note} {Our sample is non-magnetic and we use linearly polarized
  light to avoid dynamic polarization of nuclear spins.}\BibitemShut {Stop}%
\bibitem [{\citenamefont {Knox}(1963)}]{Knox}%
  \BibitemOpen
  \bibfield  {author} {\bibinfo {author} {\bibfnamefont {R.~S.}\ \bibnamefont
  {Knox}},\ }\href
  {http://www.amazon.com/exec/obidos/redirect?tag=citeulike07-20&path=ASIN/B001SA0ZZQ}
  {\emph {\bibinfo {title} {{Theory of Excitons - Supplement 5 Solid State
  Physics}}}}\ (\bibinfo  {publisher} {Academic Press},\ \bibinfo {address}
  {New York, New York},\ \bibinfo {year} {1963})\BibitemShut {NoStop}%
\bibitem [{Note2()}]{Note2}%
  \BibitemOpen
  \bibinfo {note} {Experimentally, the \({\protect \text {NA}=0.7}\) objective
  lens collects luminescence from a range of angles. In our system, light is
  collected from excitons momenta within 7\% of \(\hbar K_x\) in Eq.~\protect
  \textup {\hbox {\mathsurround \z@ \protect \normalfont (\ignorespaces \ref
  {eq:anglekagree}\unskip \@@italiccorr )}}.}\BibitemShut {Stop}%
\bibitem [{\citenamefont {Gross}\ \emph {et~al.}(1961)\citenamefont {Gross},
  \citenamefont {Zakharchenya},\ and\ \citenamefont
  {Konstantinov}}]{Gross1961}%
  \BibitemOpen
  \bibfield  {author} {\bibinfo {author} {\bibfnamefont {E.~F.}\ \bibnamefont
  {Gross}}, \bibinfo {author} {\bibfnamefont {B.~P.}\ \bibnamefont
  {Zakharchenya}}, \ and\ \bibinfo {author} {\bibfnamefont {O.~V.}\
  \bibnamefont {Konstantinov}},\ }\bibfield  {title} {\enquote {\bibinfo
  {title} {{Effect of magnetic field inversion in spectra of exciton absorption
  in CdSe crystal}},}\ }\href@noop {} {\bibfield  {journal} {\bibinfo
  {journal} {Sov. Phys. Solid State}\ }\textbf {\bibinfo {volume} {3}},\
  \bibinfo {pages} {221} (\bibinfo {year} {1961})}\BibitemShut {NoStop}%
\bibitem [{\citenamefont {Thomas}\ and\ \citenamefont
  {Hopfield}(1961)}]{Thomas1961}%
  \BibitemOpen
  \bibfield  {author} {\bibinfo {author} {\bibfnamefont {D.~G.}\ \bibnamefont
  {Thomas}}\ and\ \bibinfo {author} {\bibfnamefont {J.~J.}\ \bibnamefont
  {Hopfield}},\ }\bibfield  {title} {\enquote {\bibinfo {title} {{A
  magneto-Stark effect and exciton motion in CdS}},}\ }\href {\doibase
  10.1103/PhysRev.124.657} {\bibfield  {journal} {\bibinfo  {journal} {Phys.
  Rev.}\ }\textbf {\bibinfo {volume} {124}},\ \bibinfo {pages} {657--665}
  (\bibinfo {year} {1961})}\BibitemShut {NoStop}%
\bibitem [{\citenamefont {Jahn}\ \emph {et~al.}(2012)\citenamefont {Jahn},
  \citenamefont {L\"{a}hnemann}, \citenamefont {Pf\"{u}ller}, \citenamefont
  {Brandt}, \citenamefont {Breuer}, \citenamefont {Jenichen}, \citenamefont
  {Ramsteiner}, \citenamefont {Geelhaar},\ and\ \citenamefont
  {Riechert}}]{Jahn2012}%
  \BibitemOpen
  \bibfield  {author} {\bibinfo {author} {\bibfnamefont {U.}~\bibnamefont
  {Jahn}}, \bibinfo {author} {\bibfnamefont {J.}~\bibnamefont {L\"{a}hnemann}},
  \bibinfo {author} {\bibfnamefont {C.}~\bibnamefont {Pf\"{u}ller}}, \bibinfo
  {author} {\bibfnamefont {O.}~\bibnamefont {Brandt}}, \bibinfo {author}
  {\bibfnamefont {S.}~\bibnamefont {Breuer}}, \bibinfo {author} {\bibfnamefont
  {B.}~\bibnamefont {Jenichen}}, \bibinfo {author} {\bibfnamefont
  {M.}~\bibnamefont {Ramsteiner}}, \bibinfo {author} {\bibfnamefont
  {L.}~\bibnamefont {Geelhaar}}, \ and\ \bibinfo {author} {\bibfnamefont
  {H.}~\bibnamefont {Riechert}},\ }\bibfield  {title} {\enquote {\bibinfo
  {title} {{Luminescence of GaAs nanowires consisting of wurtzite and
  zinc-blende segments}},}\ }\href {\doibase 10.1103/physrevb.85.045323}
  {\bibfield  {journal} {\bibinfo  {journal} {Physical Review B}\ }\textbf
  {\bibinfo {volume} {85}} (\bibinfo {year} {2012}),\
  10.1103/physrevb.85.045323}\BibitemShut {NoStop}%
\bibitem [{\citenamefont {Warburton}\ \emph {et~al.}(2002)\citenamefont
  {Warburton}, \citenamefont {Schulhauser}, \citenamefont {Haft}, \citenamefont
  {Sch\"{a}flein}, \citenamefont {Karrai}, \citenamefont {Garcia},
  \citenamefont {Schoenfeld},\ and\ \citenamefont {Petroff}}]{Warburton2002}%
  \BibitemOpen
  \bibfield  {author} {\bibinfo {author} {\bibfnamefont {R.~J.}\ \bibnamefont
  {Warburton}}, \bibinfo {author} {\bibfnamefont {C.}~\bibnamefont
  {Schulhauser}}, \bibinfo {author} {\bibfnamefont {D.}~\bibnamefont {Haft}},
  \bibinfo {author} {\bibfnamefont {C.}~\bibnamefont {Sch\"{a}flein}}, \bibinfo
  {author} {\bibfnamefont {K.}~\bibnamefont {Karrai}}, \bibinfo {author}
  {\bibfnamefont {J.~M.}\ \bibnamefont {Garcia}}, \bibinfo {author}
  {\bibfnamefont {W.}~\bibnamefont {Schoenfeld}}, \ and\ \bibinfo {author}
  {\bibfnamefont {P.~M.}\ \bibnamefont {Petroff}},\ }\bibfield  {title}
  {\enquote {\bibinfo {title} {{Giant permanent dipole moments of excitons in
  semiconductor nanostructures}},}\ }\href {\doibase
  10.1103/PhysRevB.65.113303} {\bibfield  {journal} {\bibinfo  {journal} {Phys.
  Rev. B}\ }\textbf {\bibinfo {volume} {65}},\ \bibinfo {pages} {113303}
  (\bibinfo {year} {2002})}\BibitemShut {NoStop}%
\bibitem [{\citenamefont {Chai}\ \emph {et~al.}(1985)\citenamefont {Chai},
  \citenamefont {Y-c},\ and\ \citenamefont {Hierl}}]{Chai1985}%
  \BibitemOpen
  \bibfield  {author} {\bibinfo {author} {\bibfnamefont {Y.~G.}\ \bibnamefont
  {Chai}}, \bibinfo {author} {\bibnamefont {Y-c}}, \ and\ \bibinfo {author}
  {\bibfnamefont {T.}~\bibnamefont {Hierl}},\ }\bibfield  {title} {\enquote
  {\bibinfo {title} {{Elimination of ``pair" defects from GaAs layers grown by
  molecular beam epitaxy}},}\ }\href {\doibase 10.1063/1.96269} {\bibfield
  {journal} {\bibinfo  {journal} {Applied Physics Letters}\ }\textbf {\bibinfo
  {volume} {47}},\ \bibinfo {pages} {1327--1329} (\bibinfo {year}
  {1985})}\BibitemShut {NoStop}%
\bibitem [{\citenamefont {Haverkort}\ \emph {et~al.}(1992)\citenamefont
  {Haverkort}, \citenamefont {Schuwer}, \citenamefont {Leys},\ and\
  \citenamefont {Wolter}}]{Haverkort1992}%
  \BibitemOpen
  \bibfield  {author} {\bibinfo {author} {\bibfnamefont {J.~E.~M.}\
  \bibnamefont {Haverkort}}, \bibinfo {author} {\bibfnamefont {M.~P.}\
  \bibnamefont {Schuwer}}, \bibinfo {author} {\bibfnamefont {M.~R.}\
  \bibnamefont {Leys}}, \ and\ \bibinfo {author} {\bibfnamefont {J.~H.}\
  \bibnamefont {Wolter}},\ }\bibfield  {title} {\enquote {\bibinfo {title}
  {{Spatial variations of photoluminescence line broadening around oval defects
  in GaAs/AlGaAs multiple quantum wells}},}\ }\href
  {http://stacks.iop.org/0268-1242/7/i=1A/a=011} {\bibfield  {journal}
  {\bibinfo  {journal} {Semiconductor Science and Technology}\ }\textbf
  {\bibinfo {volume} {7}} (\bibinfo {year} {1992})}\BibitemShut {NoStop}%
\bibitem [{\citenamefont {Bir}\ and\ \citenamefont {Pikus}(1974)}]{BirPikus}%
  \BibitemOpen
  \bibfield  {author} {\bibinfo {author} {\bibfnamefont {G.~L.}\ \bibnamefont
  {Bir}}\ and\ \bibinfo {author} {\bibfnamefont {G.~E.}\ \bibnamefont
  {Pikus}},\ }\href
  {http://www.amazon.com/exec/obidos/redirect?tag=citeulike07-20&path=ASIN/0470073217}
  {\emph {\bibinfo {title} {{Symmetry and strain-induced effects in
  semiconductors}}}}\ (\bibinfo  {publisher} {Halsted Press},\ \bibinfo {year}
  {1974})\BibitemShut {NoStop}%
\bibitem [{\citenamefont {Koster}\ \emph {et~al.}(1963)\citenamefont {Koster},
  \citenamefont {Dimmock},\ and\ \citenamefont {Wheeler}}]{Koster1963}%
  \BibitemOpen
  \bibfield  {author} {\bibinfo {author} {\bibfnamefont {G.~F.}\ \bibnamefont
  {Koster}}, \bibinfo {author} {\bibfnamefont {J.~O.}\ \bibnamefont {Dimmock}},
  \ and\ \bibinfo {author} {\bibfnamefont {R.~G.}\ \bibnamefont {Wheeler}},\
  }\href {http://www.worldcat.org/isbn/9780262110105} {\emph {\bibinfo {title}
  {{Properties of the thirty-two point groups}}}}\ (\bibinfo  {publisher}
  {M.I.T. Press},\ \bibinfo {year} {1963})\BibitemShut {NoStop}%
\bibitem [{\citenamefont {Dresselhaus}\ \emph {et~al.}(2008)\citenamefont
  {Dresselhaus}, \citenamefont {Dresselhaus},\ and\ \citenamefont
  {Jorio}}]{Dresselhaus}%
  \BibitemOpen
  \bibfield  {author} {\bibinfo {author} {\bibfnamefont {M.~S.}\ \bibnamefont
  {Dresselhaus}}, \bibinfo {author} {\bibfnamefont {G.}~\bibnamefont
  {Dresselhaus}}, \ and\ \bibinfo {author} {\bibfnamefont {A.}~\bibnamefont
  {Jorio}},\ }\href
  {http://www.amazon.com/exec/obidos/redirect?tag=citeulike07-20&path=ASIN/3540328971}
  {\emph {\bibinfo {title} {{Group Theory: Application to the Physics of
  Condensed Matter}}}},\ \bibinfo {edition} {2008th}\ ed.\ (\bibinfo
  {publisher} {Springer},\ \bibinfo {address} {Berlin, Germany},\ \bibinfo
  {year} {2008})\BibitemShut {NoStop}%
\bibitem [{\citenamefont {Birman}(1959)}]{Birman1959}%
  \BibitemOpen
  \bibfield  {author} {\bibinfo {author} {\bibfnamefont {J.~L.}\ \bibnamefont
  {Birman}},\ }\bibfield  {title} {\enquote {\bibinfo {title} {{Some selection
  rules for band-band transitions in wurtzite structure}},}\ }\href {\doibase
  10.1103/PhysRev.114.1490} {\bibfield  {journal} {\bibinfo  {journal} {Phys.
  Rev.}\ }\textbf {\bibinfo {volume} {114}},\ \bibinfo {pages} {1490--1492}
  (\bibinfo {year} {1959})}\BibitemShut {NoStop}%
\bibitem [{\citenamefont {Dresselhaus}(1955)}]{Dresselhaus1955}%
  \BibitemOpen
  \bibfield  {author} {\bibinfo {author} {\bibfnamefont {G.}~\bibnamefont
  {Dresselhaus}},\ }\bibfield  {title} {\enquote {\bibinfo {title} {{Spin-orbit
  coupling effects in zinc blende structures}},}\ }\href {\doibase
  10.1103/PhysRev.100.580} {\bibfield  {journal} {\bibinfo  {journal} {Phys.
  Rev.}\ }\textbf {\bibinfo {volume} {100}},\ \bibinfo {pages} {580--586}
  (\bibinfo {year} {1955})}\BibitemShut {NoStop}%
\bibitem [{\citenamefont {Herring}(1937)}]{Herring1937}%
  \BibitemOpen
  \bibfield  {author} {\bibinfo {author} {\bibfnamefont {C.}~\bibnamefont
  {Herring}},\ }\bibfield  {title} {\enquote {\bibinfo {title} {{Effect of
  time-reversal symmetry on energy bands of crystals}},}\ }\href {\doibase
  10.1103/PhysRev.52.361} {\bibfield  {journal} {\bibinfo  {journal} {Phys.
  Rev.}\ }\textbf {\bibinfo {volume} {52}},\ \bibinfo {pages} {361--365}
  (\bibinfo {year} {1937})}\BibitemShut {NoStop}%
\bibitem [{\citenamefont {Elliott}(1954)}]{Elliot1954}%
  \BibitemOpen
  \bibfield  {author} {\bibinfo {author} {\bibfnamefont {R.~J.}\ \bibnamefont
  {Elliott}},\ }\bibfield  {title} {\enquote {\bibinfo {title} {{Spin-orbit
  coupling in band theory--character tables for some ``double'' space
  groups}},}\ }\href {\doibase 10.1103/PhysRev.96.280} {\bibfield  {journal}
  {\bibinfo  {journal} {Phys. Rev.}\ }\textbf {\bibinfo {volume} {96}},\
  \bibinfo {pages} {280--287} (\bibinfo {year} {1954})}\BibitemShut {NoStop}%
\bibitem [{\citenamefont {Sallen}\ \emph {et~al.}(2011)\citenamefont {Sallen},
  \citenamefont {Urbaszek}, \citenamefont {Glazov}, \citenamefont {Ivchenko},
  \citenamefont {Kuroda}, \citenamefont {Mano}, \citenamefont {Kunz},
  \citenamefont {Abbarchi}, \citenamefont {Sakoda}, \citenamefont {Lagarde},
  \citenamefont {Balocchi}, \citenamefont {Marie},\ and\ \citenamefont
  {Amand}}]{Sallen2011}%
  \BibitemOpen
  \bibfield  {author} {\bibinfo {author} {\bibfnamefont {G.}~\bibnamefont
  {Sallen}}, \bibinfo {author} {\bibfnamefont {B.}~\bibnamefont {Urbaszek}},
  \bibinfo {author} {\bibfnamefont {M.~M.}\ \bibnamefont {Glazov}}, \bibinfo
  {author} {\bibfnamefont {E.~L.}\ \bibnamefont {Ivchenko}}, \bibinfo {author}
  {\bibfnamefont {T.}~\bibnamefont {Kuroda}}, \bibinfo {author} {\bibfnamefont
  {T.}~\bibnamefont {Mano}}, \bibinfo {author} {\bibfnamefont {S.}~\bibnamefont
  {Kunz}}, \bibinfo {author} {\bibfnamefont {M.}~\bibnamefont {Abbarchi}},
  \bibinfo {author} {\bibfnamefont {K.}~\bibnamefont {Sakoda}}, \bibinfo
  {author} {\bibfnamefont {D.}~\bibnamefont {Lagarde}}, \bibinfo {author}
  {\bibfnamefont {A.}~\bibnamefont {Balocchi}}, \bibinfo {author}
  {\bibfnamefont {X.}~\bibnamefont {Marie}}, \ and\ \bibinfo {author}
  {\bibfnamefont {T.}~\bibnamefont {Amand}},\ }\bibfield  {title} {\enquote
  {\bibinfo {title} {{Dark-bright mixing of interband transitions in symmetric
  semiconductor quantum dots}},}\ }\href {\doibase
  10.1103/PhysRevLett.107.166604} {\bibfield  {journal} {\bibinfo  {journal}
  {Phys. Rev. Lett.}\ }\textbf {\bibinfo {volume} {107}},\ \bibinfo {pages}
  {166604} (\bibinfo {year} {2011})}\BibitemShut {NoStop}%
\bibitem [{\citenamefont {Burns}(1977)}]{Burns}%
  \BibitemOpen
  \bibfield  {author} {\bibinfo {author} {\bibfnamefont {G.}~\bibnamefont
  {Burns}},\ }\href
  {http://www.amazon.com/exec/obidos/redirect?tag=citeulike07-20&path=ASIN/1483175685}
  {\emph {\bibinfo {title} {{Introduction to group theory with applications:
  materials science and technology}}}}\ (\bibinfo  {publisher} {Academic
  Press},\ \bibinfo {year} {1977})\BibitemShut {NoStop}%
\bibitem [{\citenamefont {Ivchenko}\ and\ \citenamefont
  {Pikus}(1997)}]{Ivchenko1997}%
  \BibitemOpen
  \bibfield  {author} {\bibinfo {author} {\bibfnamefont {E.~L.}\ \bibnamefont
  {Ivchenko}}\ and\ \bibinfo {author} {\bibfnamefont {G.~E.}\ \bibnamefont
  {Pikus}},\ }\href {http://www.worldcat.org/isbn/9783642644931} {\emph
  {\bibinfo {title} {{Superlattices and other heterostructures: symmetry and
  optical phenomena}}}}\ (\bibinfo  {publisher} {Springer},\ \bibinfo {year}
  {1997})\BibitemShut {NoStop}%
\bibitem [{\citenamefont {Ikonic´}\ \emph {et~al.}(1992)\citenamefont
  {Ikonic´}, \citenamefont {Milanovic´},\ and\ \citenamefont
  {Tjapkin}}]{Ikonic1992}%
  \BibitemOpen
  \bibfield  {author} {\bibinfo {author} {\bibfnamefont {Z.}~\bibnamefont
  {Ikonic´}}, \bibinfo {author} {\bibfnamefont {V.}~\bibnamefont
  {Milanovic´}}, \ and\ \bibinfo {author} {\bibfnamefont {D.}~\bibnamefont
  {Tjapkin}},\ }\bibfield  {title} {\enquote {\bibinfo {title} {{Valence
  subband structure of [100]-, [110]-, and [111]-grown GaAs-(Al,Ga)As quantum
  wells and the accuracy of the axial approximation}},}\ }\href {\doibase
  10.1103/physrevb.46.4285} {\bibfield  {journal} {\bibinfo  {journal}
  {Physical Review B}\ }\textbf {\bibinfo {volume} {46}},\ \bibinfo {pages}
  {4285--4288} (\bibinfo {year} {1992})}\BibitemShut {NoStop}%
\end{thebibliography}%

\balancecolsandclearpage
\end{document}